\documentclass{jfm}
\usepackage{amsfonts}
\usepackage{amssymb}
\usepackage{amsbsy}
\usepackage{amsmath}
\usepackage{graphicx}
\usepackage{float}
\usepackage{stfloats}
\usepackage{color}
\usepackage{subfig}
\usepackage{array}

\ifCUPmtlplainloaded \else
  \checkfont{eurm10}
  \iffontfound
    \IfFileExists{upmath.sty}
      {\typeout{^^JFound AMS Euler Roman fonts on the system,
                   using the 'upmath' package.^^J}%
       \usepackage{upmath}}
      {\typeout{^^JFound AMS Euler Roman fonts on the system, but you
                   dont seem to have the}%
       \typeout{'upmath' package installed. JFM.cls can take advantage
                 of these fonts,^^Jif you use 'upmath' package.^^J}%
      }
  \else
  \fi
\fi

\ifCUPmtlplainloaded \else
  \checkfont{msam10}
  \iffontfound
    \IfFileExists{amssymb.sty}
      {\typeout{^^JFound AMS Symbol fonts on the system, using the
                'amssymb' package.^^J}%
       \usepackage{amssymb}%
         \let\leq=\leqslant
         \let\geq=\geqslant
      }{}
  \fi
\fi

\ifCUPmtlplainloaded \else
  \IfFileExists{amsbsy.sty}
    {\typeout{^^JFound the 'amsbsy' package on the system, using it.^^J}%
     \usepackage{amsbsy}}
    {\providecommand\boldsymbol[1]{\mbox{\boldmath $##1$}}}
\fi

\begin{document}

\shorttitle{Shear induced migration of microswimmers in a pressure driven flow} 
\shortauthor{LaxminarsimhaRao V. et al} 

\title{Shear induced migration of microswimmers in pressure-driven channel flow }
\graphicspath{{Figures_InhomSpat/}}

\author
 {
 LaxminarsimhaRao V.,
  Sankalp Nambiar
  \and 
  Ganesh Subramanian
  \corresp{\email{sganesh@jncasr.ac.in}}
  }

\affiliation
{
Engineering Mechanics Unit, Jawaharlal Nehru Centre for Advanced Scientific Research Bnagalore, India
}

\maketitle

\begin{abstract}
We study the shear induced migration of microswimmers (primarily, active Brownian particles or ABP's) in plane Poiseuille flow. For wide channels characterized by $U_s/HD_r \ll 1$, the separation between time scales characterizing the swimmer orientation dynamics (of O($D^{-1}_r$)) and those that characterize migration across the channel (of  O($H^{2}D_r/U^{2}_s$)), allows for use of the method of multiple scales to derive a drift-diffusion equation for the swimmer concentration profile; here, $U_s$ is the swimming speed, $H$ is the channel half-width, and $D_r$ is the swimmer rotary diffusivity. The steady state concentration profile is a function of the P\'eclet number, $Pe = U_{f}/(D_r H)$ ($U_f$ being the channel centerline velocity), and the swimmer aspect ratio $\kappa$. Swimmers with $ \kappa \gg 1$ (with $ \kappa \sim$ O(1)), in the regime $1 \ll \textit{Pe} \ll \kappa^3$ ($Pe\sim$ O(1)), migrate towards the channel walls, corresponding to a high-shear trapping behavior. For  $Pe \gg \kappa^3 $ ($Pe \gg $ 1 for $\kappa \sim$ O(1)), however, swimmers migrate towards the centerline, corresponding to a low-shear trapping behavior. Interestingly, within the low-shear trapping regime, swimmers with $\kappa < 2$ asymptote to a $Pe$-independent concentration profile for large $Pe$, while those with $\kappa \geq 2$ exhibit  a `centerline-collapse' for $Pe \to \infty$. The prediction of low-shear-trapping, validated by Langevin simulations, is the first explanation of recent experimental observations [\cite{barry2015shear}]. We organize the high-shear and low-shear trapping regimes on a $Pe-\kappa$ plane, thereby highlighting the singular behavior of infinite-aspect-ratio swimmers.  

\end{abstract}

\section{Introduction} \label{sec:introduction}
Biologically active microswimmers, a subset of the low-Reynolds-number dwellers, have motivated research spanning over several decades [\cite{hancock1953self, gray1955propulsion, brennen1977fluid, pedley1992hydrodynamic, koch2011collective, guasto2012fluid, marchetti2013hydrodynamics}]. Owing to their intrinsic activity, suspensions of such microswimmers exhibit a host of interesting phenomena such as enhanced tracer diffusion [\cite{Wu2000, underhill2008diffusion, leptos2009dynamics, deepak2015, arratia2016, stenhammar2017role}], long-ranged orientational order [\cite{saintillan2007orientational, saintillan2008instabilities, underhill2011correlations, stenhammar2017role,nambiar2019enhanced}], negative viscosities [\cite{lopez2015turning, bechtel2017linear, nambiar2017stress,nambiar2018stress,takatori2014swim1,takatori2014swim,takatori2015towards}], collective motion/bacterial turbulence [\cite{dombrowski2004self, underhill2008diffusion, subramanian2009critical, subramanian2011stability, wensink2012meso, marchetti2013hydrodynamics}], among others. The underlying mechanisms driving most of these phenomena may be understood from studying quiescent swimmer suspensions, or those subjected to homogeneous shear flows. However, nearly all biologically relevant problems involve motile microorganisms in complex flow environments [\cite{guasto2012fluid, rusconi2014microfluidics}]. Complex flows, with or without boundaries, are known to alter the nutrient landscape [\cite{taylor2012trade}] influence swimmer rheotaxis [\cite{stocker2006microorganisms, fu2012bacterial}], result in shear induced migration [\cite{rusconi2014bacterial, bearon2015trapping, barry2015shear}], in turn leading to shear-banding instabilities [\cite{guo2018symmetric,laxman2018}], and possibly, influencing biofilm formation [\cite{rusconi2010laminar, rusconi2014microfluidics, kim2014filaments}]. The study in this paper examines the orientation dynamics and transport of microorganisms in one of the simplest inhomogeneous shearing flows - the pressure-driven flow between a pair of parallel plates. 

Microorganisms, bacteria such as \emph{E. coli} and \emph{B. subtilis}, algae such as \emph{C. reinhardtii} [\cite{berg2008coli, elgeti2015physics, guasto2012fluid}], and even artificial swimmers, exhibit an intrinsic stochasticity in their swimming motion owing to a tendency to reorient as they swim. The intrinsic stochasticity may be biologically motivated [\cite{berg2008coli, elgeti2015physics, guasto2012fluid}], or a consequence of reaction with the solvent [\cite{ebbens2014electrokinetic, moran2017phoretic}] or driven by an external field [\cite{chaturvedi2010magnetic, fischer2011magnetically, peyer2013bio}]. The reorientations occur either as discrete finite-amplitude events leading to run-and-tumble dynamics, or as continuous small-amplitude events leading to active Brownian dynamics. Regardless of its origin or detailed nature, the interplay between the stochastic dynamics of the swimmer and an imposed inhomogeneous shearing flow has been shown to result in migration patterns [\cite{chilukuri2014impact,rusconi2014bacterial,chilukuri2015dispersion,barry2015shear,ezhilan2015transport,bearon2015trapping, sokolov2016rapid,sokolov2018instability}] that stand in sharp contrast to those known for suspensions of passive particles [\cite{gadala1980shear, leighton1987shear, koh1994experimental, nott1994pressure, nitsche1997shear, strednak2018shear}]. Experiments and simulations on suspensions of passive particles in pressure-driven channel flow reveal, irrespective of particle geometry (rigid spheres and/or slender fibers), migration towards the channel center [\cite{koh1994experimental, strednak2018shear, nott1994pressure}]. Microswimmers on the other hand, depending on swimmer geometry and flow characteristics, exhibit migration both towards the wall (high-shear trapping) and channel center (low-shear trapping)  [\cite{rusconi2014bacterial,barry2015shear,bearon2015trapping}]. Recent experiments in a  pressure-driven microchannel flow have found slender bacteria to migrate towards the walls [\cite{rusconi2014bacterial}], whereas relatively round algae have been found to migrate towards the wall or the centerline  depending on the particular species and the prevailing shear rate [\cite{barry2015shear}]. Earlier computations by \cite{ezhilan2015transport}, for infinitely slender swimmers have only confirmed the existence of high-shear trapping over a range of $Pe$. Here, $Pe = U_f/(D_r H)$ is the rotary P\'eclet number, the shear rate ($U_f/H$) measured in units of the inverse rotary diffusivity ($D^{-1}_r$); $U_f$ and $H$ are, respectively, the centerline velocity and channel half-width, and $D_r$ the swimmer diffusivity. An independent computational investigation by \cite{bearon2015trapping} revealed regimes of both high and low-shear trapping, but this was in contrast to the authors' analysis which, similar to \cite{ezhilan2015transport}, predicted high-shear trapping regardless of swimmer aspect ratio ($\kappa$); this apparent discrepancy remains unexplained. There have been additional simulations [\cite{chilukuri2014impact,chilukuri2015dispersion}] examining the spatial distribution, and the dispersion in the flow direction, of microswimmers (modeled as dumbbells), in pressure driven channel flow. The channel in these cases was, however, quite narrow, and as a result, confinement effects, together with wall-mediated hydrodynamic interactions were dominant. In contrast, the experiments mentioned above used  channels much wider than the single swimmer dimension.
 
From the above summary, it is clear that a comprehensive understanding of the interplay between swimmer shape and shear induced migration in a channel flow is lacking. In particular, there has been no attempt to interpret and explain the observations of \cite{rusconi2014bacterial} and \cite{barry2015shear} based on a common theoretical frame work. In this work, we use both theory and Langevin simulations, to comprehensively characterize shear induced migration of microswimmers in the $Pe-\kappa$ plane, rationalizing the aforementioned observations in the process. Two distinct classes of swimmer concentration profiles are identified in this parameter plane, corresponding to high-shear and low-shear trapping behavior. In \textsection\ref{theoretical_model}, the theoretical framework and the simulation protocol adopted are described. In \textsection\ref{subsec:conc}, beginning with the equation governing the swimmer probability density in position-orientation space, and using the separation of the fast (O$(D_r^{-1})$) reorientation time scale and the much slower (O$(H^2D_r/U_s^2)$) time scale for migration across the channel ($U_s$ here being the swimming speed), we derive a drift-diffusion equation for the cross-stream swimmer concentration profile with the aid of the method of multiple scales. This is followed by the description of the simulation scheme in \textsection\ref{subsec:Langevin_sim}; the scheme obtains the concentration profile by numerically integrating the Langevin equations of motion for the individual swimmers with periodic boundary conditions. The steady state swimmer concentration profiles obtained from the multiple scales analysis are presented in \textsection\ref{Sec:4}; high-shear trapping is discussed in \textsection\ref{Sec:3.1} and low-shear trapping in \textsection\ref{Sec:3.2}. In each case comparisons are drawn with both experiments [\cite{rusconi2014bacterial, barry2015shear}], and the Langevin simulations described in \textsection\ref{subsec:Langevin_sim}. In  \textsection\ref{sec:depletion_index}, the swimmer depletion index, a measure of the spatial inhomogeneity of the swimmer concentration, is plotted as a function of \textit{Pe} for swimmers of different aspect ratios, highlighting both the singular behavior of infinitely slender swimmers, and the centerline-collapse that occurs for swimmers with $\kappa$ finite but greater than (approximately) 2, for $Pe \rightarrow \infty$, in the low-shear trapping regime. Next, the shear induced migration behavior is organized on the $Pe-\kappa$ plane, demarcating the low and high-shear trapping regimes. Finally, in \textsection\ref{Conclusion}, we present concluding remarks, that include scaling arguments for the threshold governing the transition from active to passive shear induced migration patterns, and directions for future work.                

\section{Theoretical framework and the Langevin simulation protocol} \label{theoretical_model}
In \textsection\ref{subsec:conc}, we use the method of multiple scales to derive the swimmer concentration profiles as a function of the transverse coordinate in plane Poiseuille flow. Next, in \textsection\ref{subsec:Langevin_sim}, we describe the scheme adopted to simulate a discrete system of swimmers in the same flow.
\begin{figure}
\center
\includegraphics[scale=.7]{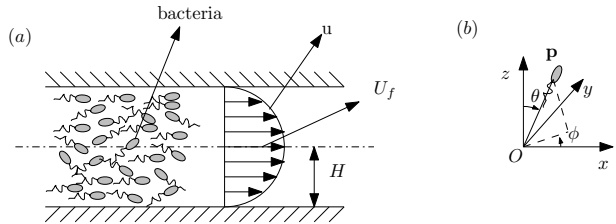}
\caption{Schematic showing (a) a dilute swimmer (bacterial) suspension subject to a parabolic shearing flow ($u$) in a channel of width $2H$, and with a centerline velocity $U_f$. (b) The coordinate system used analyze the swimmer orientation distributions.}
\label{FIG:Schematic}
\end{figure}

\subsection{Theoretical framework: the method of multiple scales} \label{subsec:conc}
The time evolution of the probability density $\Omega(\mathbf{x}, \mathbf{p}, t)$ for a dilute swimmer suspension subject to a shearing flow (see figure \ref{FIG:Schematic}), is given by [\cite{subramanian2009critical}]: 

\begin{equation}
\frac{\partial \Omega
}{\partial t}+U_{s}\nabla_{\mathbf{x}}\cdot(\Omega \mathbf{p})+\nabla_{p}\cdot(\dot{\mathbf{p}}\Omega)+\frac{1}{\tau}\left[\Omega-\int d \mathbf{p}^{\prime} K(\mathbf{p}|\mathbf{p}^{\prime})\Omega(\mathbf{p}^{\prime}) \right]-D_{r}\nabla^{2}_{p}\Omega=0,
\label{EQ:GOvEq}
\end{equation}
where $\mathbf{x}$ and $\mathbf{p}$ denote the swimmer position and orientation. The second term on the left hand side of (\ref{EQ:GOvEq}) denotes spatial convection of the probability density owing to swimming with speed $U_s$, and the term involving $\dot{\mathbf{p}}$ denotes swimmer rotation by the ambient shear. The terms within brackets model a run-and-tumble process [\cite{berg1993random, subramanian2009critical, nambiar2017stress}], obeying Poission statistics with a mean run duration $\tau$; the kernel  $K(\mathbf{p}|\mathbf{p}^{\prime})$ characterizes correlation between the pre ($\mathbf{p}^{\prime}$)- and post ($\mathbf{p}$)-tumble orientations, with $K(\mathbf{p}|\mathbf{p}^{\prime})=1/(4\pi)$ for random tumbles. The last term on the left side of (\ref{EQ:GOvEq}) models the stochastic orientation change due to rotary diffusion, $D_r$ being the rotary diffusivity.

In the experiments of \cite{rusconi2014bacterial} and \cite{barry2015shear}, the ratio of the channel width ($H$) to the swimmer size ($L$) is approximately $40$. Thus, it is reasonable to assume that the swimmer senses a local linear flow, with its rotation governed by the Jeffery equation; therefore, $\dot{\mathbf{p}}=B \left[ \mathbf{E}\cdot \mathbf{p}-\mathbf{p}(\mathbf{E}:\mathbf{p}\mathbf{p})\right]+\mathbf{W}\cdot\mathbf{p}$. Here, $B=(\kappa^{2}-1)/(\kappa^{2}+1)$ is the Bretherton constant, $E_{ij}=\dot{\gamma}(z)(\delta_{i1}\delta_{j3}+\delta_{i3}\delta_{j1})/2$ and $W_{ij}=\dot{\gamma}(z)(\delta_{i1}\delta_{j3}-\delta_{i3}\delta_{j1})/2$ are the strain rate and vorticity tensors, with $\dot{\gamma}(z)=du/dz$ being the shear rate at $z$, and $u(z)=U_f(1-(z/H)^2)$ for the plane Poiseuille flow under consideration. In the above, $\delta_{ij}$ is the Kronecker delta, and the coordinate system used appears in figure \ref{FIG:Schematic}. In using Jeffery's equation above, we model the swimmers as equivalent spheroids with an aspect ratio $\kappa$ [\cite{leal1971effect}], and it is via $\kappa$ that the swimmer geometry enters the analysis, determining the nature of the shear induced migration. This implicitly assumes an axisymmetric cross-section, which is a reasonable assumption, at least for bacteria [\cite{darnton2007torque, das2018computing}], on account of the rapid (counter)-rotation on time scales much shorter than those characterizing shear induced migration.
 
Non-dimensionalizing (\ref{EQ:GOvEq}) using $D_r^{-1}$, $U_{f}$, and $H$ as scales for time, velocity and length, respectively, yields:
 \begin{equation}
\frac{\partial \Omega
}{\partial t}+\epsilon \nabla_{\mathbf{x}}\cdot(\Omega \mathbf{p})+Pe \dot{\gamma} \nabla_{p}\cdot(\tilde{\dot{\mathbf{p}}}\Omega)+\frac{1}{\tau D_{r}}\left[\Omega-\int d \mathbf{p}^{\prime} K(\mathbf{p}/\mathbf{p}^{\prime})\Omega(\mathbf{p}^{\prime}) \right]-\nabla^{2}_{p}\Omega=0,
\label{EQ:NonDimGOvEq}
\end{equation}
where $\tilde{\dot{\mathbf{p}}}=\dot{\mathbf{p}}/\dot{\gamma}(z)$ is a non-dimensional rotation rate. In (\ref{EQ:NonDimGOvEq}), $\epsilon$, \textit{Pe} and $\kappa$ are dimensionless parameters that primarily determine the different regimes of shear induced migration. The parameter $\epsilon=U_{s}/D_rH$, defined as the ratio of the swimmer mean free path $U_s/D_r$ to $H$, may be regarded as a swimmer Knudsen number. Note that in earlier efforts [\cite{takatori2014swim,takatori2017superfluid}], $\epsilon^{-1}$, which may be written as $U_s H/D_t$, with $D_t\sim U_s^2/(6 D_r)$ being the translational diffusivity, has been interpreted as the swim P\'eclet number. The parameter $\tau D_r$ determines the roles of tumbling vis-a-vis rotary diffusion in the swimmer orientation dynamics. Active Brownian particles (ABP's) correspond to $\tau D_r\rightarrow \infty$, and run-and-tumble particles (RTP's) to $\tau D_r \rightarrow 0$ [\cite{tailleur2009sedimentation, saintillan2010extensional}]; in the latter case, the swimmer Knudsen number is defined as $\epsilon=U\tau/H$ instead. Regardless of the particular value of $\tau D_r$, swimmers sample the channel cross-section diffusively for long times, the translational diffusivity being given by  $D_t=(U^2_s/(6D_r))(\tau D_r/(0.5+\tau D_r))$ [\cite{berg1993random, koch2011collective, subramanian2012fluid}] in the general case; for $\tau D_r\to\infty$, $D_t=U^2_{s}/(6D_r)$ as above. In the experiments of \cite{rusconi2014bacterial}, $\epsilon\sim 0.23$ for the bacterium {\textit{B. subtilis}}, and in \cite{barry2015shear},  $\epsilon\sim 0.29$ and 0.15 for the algal species \textit{Heterosigma} and \textit{Dunaliella}, respectively. For these small values of $\epsilon$, the time scales characterizing the swimmer orientation dynamics (of O($\tau$) or O($D^{-1}_r$)) and the diffusive sampling of the channel cross-section (of O($H^{2}/U^{2}_{s}\tau$) or  O($H^{2}D_r/U^{2}_{s}$)) are well separated, their ratio being of O($\epsilon^2$). Further, since the experimental range of flow rates correspond to $U_s\ll U_f (\epsilon \ll Pe)$, we use the method of multiple scales to analyze the swimmer concentration profiles [\cite{subramanian2004multiple,nitsche1997shear,kasyap2014instability,kasyap2012chemotaxis,laxman2018}] for small $\epsilon$, but with $Pe$ and $\kappa$ being arbitrary. \\

As a first step, we split the time derivative in (\ref{EQ:NonDimGOvEq}) in terms of the fast intrinsic time scale $t_{1}=t$ and the slow diffusive time scale $t_2=\epsilon^{2} t$, resulting in $\epsilon$:
 \begin{equation}
\frac{\partial \Omega
}{\partial t_1} + \epsilon^2\frac{\partial \Omega
}{\partial t_2} +\epsilon \nabla_{\mathbf{x}}\cdot(\Omega \mathbf{p})+Pe_z \nabla_{p}\cdot( \tilde{\dot{\mathbf{p}}}\Omega)+\frac{1}{\tau D_{r}}\left[\Omega-\int d \mathbf{p}^{\prime} K(\mathbf{p}/\mathbf{p}^{\prime})\Omega(\mathbf{p}^{\prime}) \right]-\nabla^{2}_{p}\Omega=0,
\label{EQ:NonDimGOvEq_MTS}
\end{equation}
where $Pe_z=Pe\dot{\gamma}(z)$ is the local P\'eclet number. Now, expanding $\Omega$ as $\Omega(z,\mathbf{p},t;\epsilon) = \Omega_{0}(z,\mathbf{p},t_1,t_2)+\epsilon \Omega_{1}(z,\mathbf{p},t_1,t_2)+\epsilon^{2} \Omega_{2}(z,\mathbf{p},t_1,t_2)+\ldots$, yields the following set of equations at successive orders in $\epsilon$:
\begin{eqnarray}
\label{EQ:Omega0}
\text{O}(1)&:&Pe_z\ \nabla_{p}\cdot(\tilde{\dot{\mathbf{p}}}G)+\frac{1}{\tau D_{r}}\left[G-\int d \mathbf{p}^{\prime} K(\mathbf{p}/\mathbf{p}^{\prime})G(\mathbf{p}^{\prime}) \right]-\nabla^{2}_{p}G=0, \\
\label{EQ:Omega1} \text{O}(\epsilon)&:&  Pe_z \nabla_{p}\cdot(\tilde{\dot{\mathbf{p}}}\Omega_{1})+\frac{1}{\tau D_{r}}\left[\Omega_{1}-\int d \mathbf{p}^{\prime} K(\mathbf{p}/\mathbf{p}^{\prime})\Omega_{1}(\mathbf{p}^{\prime}) \right]-\nabla^{2}_{p}\Omega_{1}=-\nabla_{\mathbf{x}}\cdot(\Omega_{0} \mathbf{p}),\\
\label{EQ:Omega2}\text{O}(\epsilon^2)&:& Pe_z \nabla_{p}\cdot(\tilde{\dot{\mathbf{p}}}\Omega_{2})+\frac{1}{\tau D_{r}}\left[\Omega_{2}-\int d \mathbf{p}^{\prime} K(\mathbf{p}/\mathbf{p}^{\prime})\Omega_{2}(\mathbf{p}^{\prime}) \right]-\nabla^{2}_{p}\Omega_{2}=-\frac{\partial \Omega_{0}}{\partial t_{2}}-\nabla_{\mathbf{x}}\cdot(\Omega_{1} \mathbf{p}).\nonumber\\
\end{eqnarray}
Here, we have written $\Omega_0$ as: $\Omega_0 = F(z, t_2)G(\mathbf{p}; Pe_z)$, with $\int d \mathbf{p} \Omega(z,\mathbf{p},t_{2})=F(z,t_{2})$ denoting the $z$-dependent (normalized)  swimmer concentration profile. The aforementioned ansatz assumes a quasi-steady response ($t_1 \rightarrow \infty$), with the swimmer orientation distribution having relaxed to a $Pe_z$-dependent local equilibrium; accordingly, the fast time derivative ($\partial/\partial t_1$) has been omitted in (\ref{EQ:Omega0}-\ref{EQ:Omega2}) . Writing $\Omega_1$ in (\ref{EQ:Omega1}) as $\Omega_{11}\partial F/\partial z+\Omega_{12} F$ on account of linearity, and using this form in (\ref{EQ:Omega2}) and integrating over orientation space, leads to a drift-diffusion equation for $F$:
\begin{equation}
\frac{\partial F(z,t_2)}{\partial t_{2}}=\frac{\partial}{\partial z}\left( D_{zz}(z;Pe,\kappa) \frac{\partial F(z,t_2)}{\partial z}-V_{z}(z;Pe,\kappa)F(z,t_{2}) \right),
\label{EQ:DriftDiffEquation}
\end{equation}
where the diffusivity $D_{zz}=-\int d \mathbf{p} \cos \theta \Omega_{11}$ and drift $V_z=\int d \mathbf{p} \cos \theta \Omega_{12}$. At steady state, for impermeable walls:
\begin{equation}
F(z)=\aleph \exp\left(\int_{-1}^{z} dz^{\prime} \frac{V_z(z^{\prime})}{D_{zz}(z^{\prime})}  \right),
\label{EQ:SSConcProf}
\end{equation}
where \[\aleph=\frac{1} {\int_{-1}^{1} dz \exp\left(\int_{-1}^{z} dz^{\prime} \frac{V_z(z^{\prime})}{D_{zz}(z^{\prime})}  \right)  }\]
is determined from the normalization condition $\int_{-1}^{1}dz F(z)=1$. Determining $D_{zz}$ and $V_{z}$ requires knowledge of the orientation distribution as a function of $Pe$ and $\kappa$. For arbitrary $Pe$, the orientation distributions, at different orders in $\epsilon$, are written as a truncated series in spherical harmonics, with the resulting coefficients being obtained numerically [\cite{nambiar2019enhanced}] (see appendix \ref{App:FinitePe}), while for small $Pe$, the orientation distribution is determined analytically via a perturbation expansion (see appendix \ref{App:SmallPe}).

\subsection{Langevin simulation scheme} \label{subsec:Langevin_sim}
The objective is to simulate a population of $N$ swimmers subject to a parabolic flow profile, with the purpose of determining the cross-stream swimmer concentration profile. Thus, there are three degrees of freedom per swimmer: two orientation degrees of freedom and one spatial (the swimmer $z$ coordinate). In other words, while we keep track of the swimmer orientation in three dimensions, only the cross-stream location $z$ of each swimmer is recorded. Following the experimental protocol of \cite{rusconi2014bacterial}, the swimmers are uniformly distributed along $z$ at the initial time, with their orientations following an isotropic distribution. Thereafter, at each time step $\delta t$, the swimmer orientation changes on account of either rotary diffusion or run-and-tumble motion, and due to the rotation by the flow. Here, the rotary diffusion process is modeled using the method outlined in \cite{ghosh2012gaussian}, whereas, run-and-tumble motion is implemented in accordance with Poisson statistics [\cite{berg1993random, deepak2015}]. The flow-induced rotation at any $z$ is determined using (\ref{EQ:NonDimGOvEq}). Following the update in the swimmer orientations, their positions are updated using an $RK2^{nd}$ order scheme, while also accounting for periodic boundary conditions in the $z$-direction. To avoid artefacts arising from a jump in shear rate which the swimmer encounters while swimming across the cross-stream (periodic) boundaries, the flow profile is chosen to be a double Poiseuille flow, which implies that the spatial extent of the simulation domain is $z\in [0, 4H]$. Nevertheless, we have verified that both single and double Poiseuille flow profiles yield the same steady state concentration profiles. Every result reported here is obtained from simulating a set of ten independent initial conditions, with a given run having 18000 swimmers; every run has been averaged over each of the four half-widths ($H$) of the double Poiseuille flow profile. For most runs, we fix $\epsilon = 0.05$, and find this to be sufficient to capture the $\epsilon\ll 1$ regime that pertains to the multiple scales analysis. 

\section{Results and Discussion}\label{Sec:4}
Herein, the results for the steady state cross-stream swimmer concentration profiles are presented primarily for ABP's. While RTP's exhibit analogous behavior for small to O(1) $Pe$'s, the location of the boundary separating the high-shear and low-shear trapping regimes differs. The resulting implications, and the associated sensitivity of this boundary to the precise swimming mechanism, will be reported in a later communication. We first discuss swimmer migration towards walls leading to high-shear trapping, highlighting the ranges of $\kappa$ and $Pe$ that correspond to this behavior (\textsection\ref{Sec:3.1}), before moving on to examining migration towards the centerline leading to low-shear trapping (\textsection\ref{Sec:3.2}).

\subsection{Swimmer migration towards walls - High-shear trapping} \label{Sec:3.1}

\begin{figure}
 \centering
\includegraphics[width=1\textwidth]{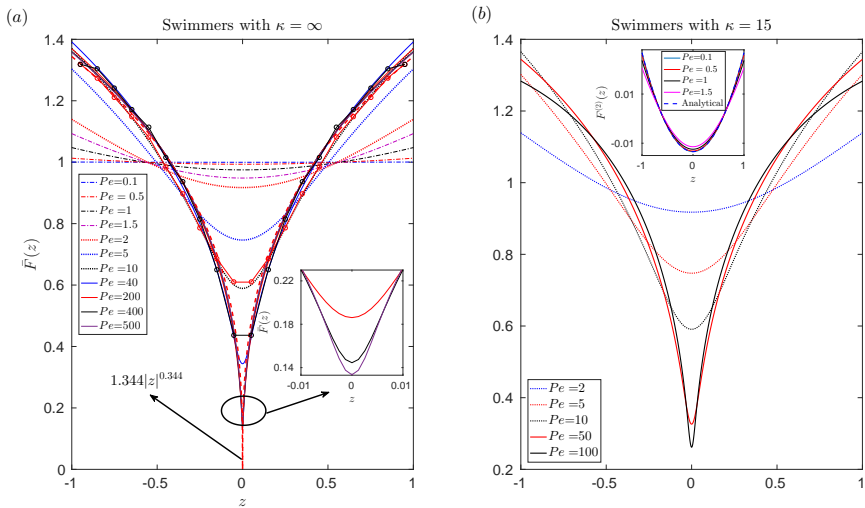}
  \caption{Swimmer concentration profiles for different \textit{Pe}: (a) $\kappa=\infty$, (b) $\kappa=15$. The red-dashed line in (a) denotes the $Pe \rightarrow \infty$ asymptote; the red and black lines with circles denote the simulation results for $Pe=10$ and 40, respectively. The inset in (a) highlights the concentration profiles, at the three largest \textit{Pe}'s, in the vicinity of the channel centerline. The inset in (b) is the comparison of the numerical profiles for $F^{(2)}(z) = (F - F^{(0)})/Pe^2$, with the parabolic depletion predicted by the small-$Pe$ analysis.}
  \label{FIG:1}
\end{figure}

\begin{figure}
\center
\includegraphics[width=1 \textwidth]{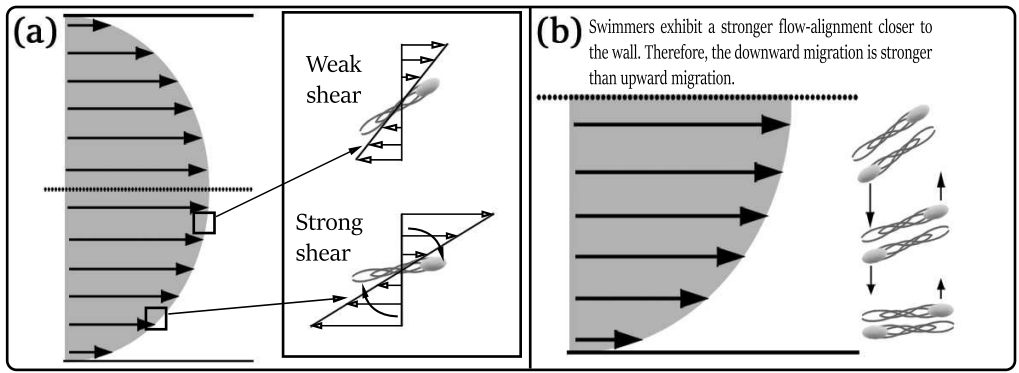}
\caption{A schematic highlighting the physical mechanism underlying the wallward migration in the high-shear trapping regime.}
\label{FIG:HighPeSchematic}
\end{figure}

In figure \ref{FIG:1}, we plot the normalized steady state concentration profile [$\bar{F}(z)=F(z, t_2\to\infty)/F(z, 0)$], using (\ref{EQ:SSConcProf}) with $D_{zz}$ and $V_z$ determined numerically as mentioned in \textsection\ref{subsec:conc}, for infinitely slender swimmers ($\kappa=\infty$), and for $\kappa = 15$, over a range of $Pe$. Also shown are the profiles obtained from Langevin simulations, for $\kappa=\infty$, for $Pe=10$ and 40, which are in good agreement with the predictions of the multiple scales analysis. All profiles exhibit near-center depletion caused by wallward migration of swimmers. For small $Pe$, a power series expansion of the swimmer probability density in $Pe$, of the form $F(z)=F^{(0)}+Pe^{2}F^{(2)}(z)+\ldots$, yields a parabolic depletion profile at O($Pe^2$), the quadratic scaling being consistent with invariance to flow reversal (see appendix \ref{App:SmallPe}). As shown in the inset of figure \ref{FIG:1}b, for $\kappa=15$ and $Pe\leq 0.5$, the numerical profile for $F^{(2)}(z)=(F-F^{(0)})/Pe^2$ agrees well with the small-$Pe$ analytical solution. With increasing $Pe$, there is a qualitative change in the shape of the concentration profiles (the profile curvature changing sign as one moves from the centerline towards the walls), and for $\kappa=\infty$, the profiles appear to asymptote to a singular cusped profile in the limit $Pe\rightarrow \infty$ (see figure \ref{FIG:1}a). The mechanism underlying near-center depletion is readily explained. For small $\epsilon$, the swimmer orientation distribution rapidly attains a $Pe$-dependent equilibrium at each $z$, with the peak of this distribution aligning with the flow direction with increasing $Pe$. Since the local P\'eclet number, $Pe_z$, increases away from the centerline (where $Pe_z=0$) as shown in figures \ref{FIG:HighPeSchematic}a and b, swimmers are, on average, more flow-aligned near the walls. The resulting reduction in the gradient component of the swimming velocity leads to a net wallward migration, eventually leading to an inhomogeneous steady state where this migration is balanced by an opposing diffusive flux.   

\begin{figure}
 \centering
\includegraphics[width=1\textwidth]{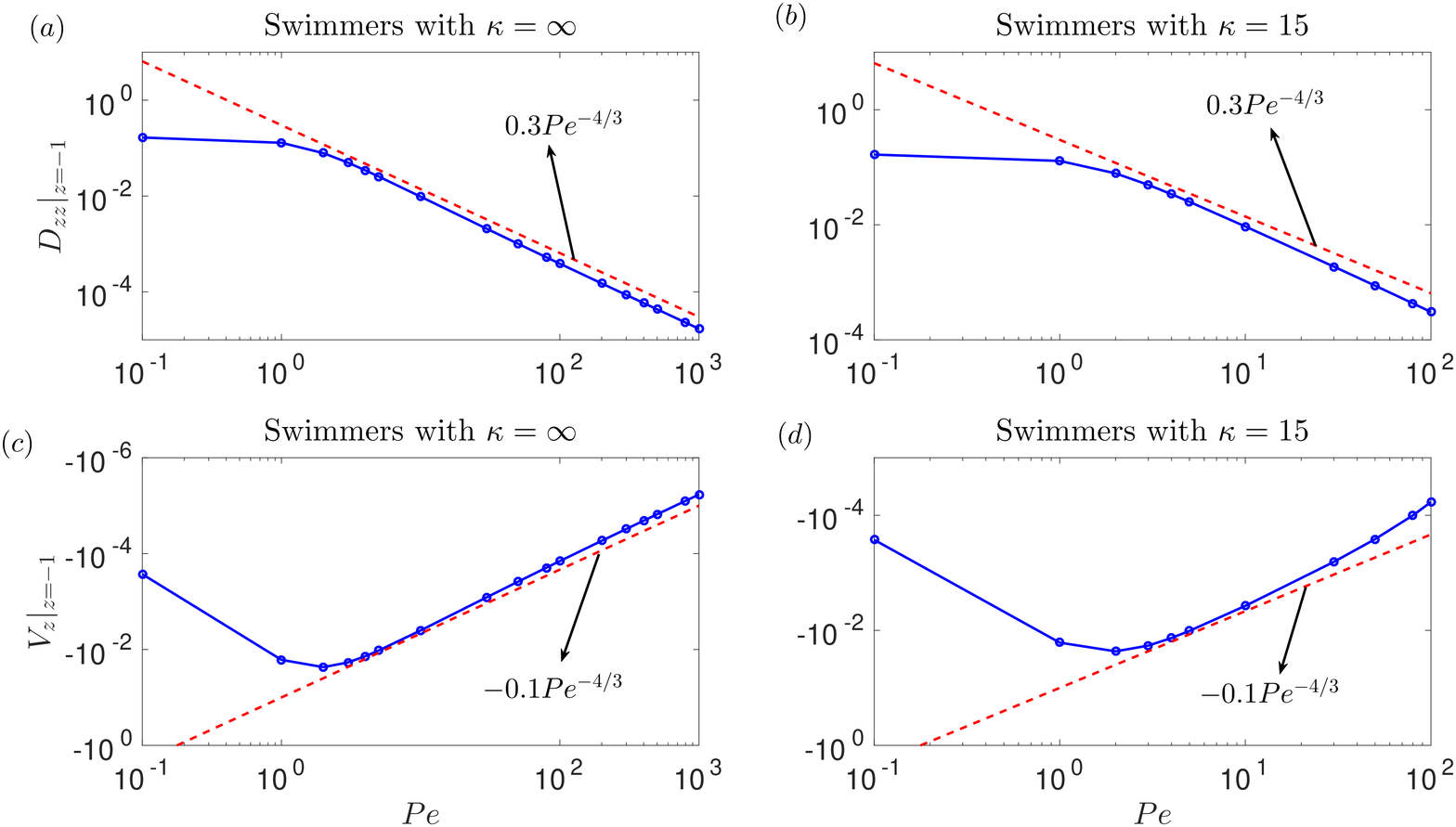}
  \caption{The swimmer diffusivity at the wall $D_{zz}\vert_{z=-1}$, as a function of $Pe$, for swimmers with $\kappa=\infty$ ($a$), $\kappa=15$ ($b$). The variation of swimmer drift at the wall $V_{z}\vert_{z=-1}$, against $Pe$, for swimmers with $\kappa=\infty$ ($c$), $\kappa=15$ ($d$).}  
  \label{FIG:11}
\end{figure}

\begin{figure}
\includegraphics[width=1\textwidth]{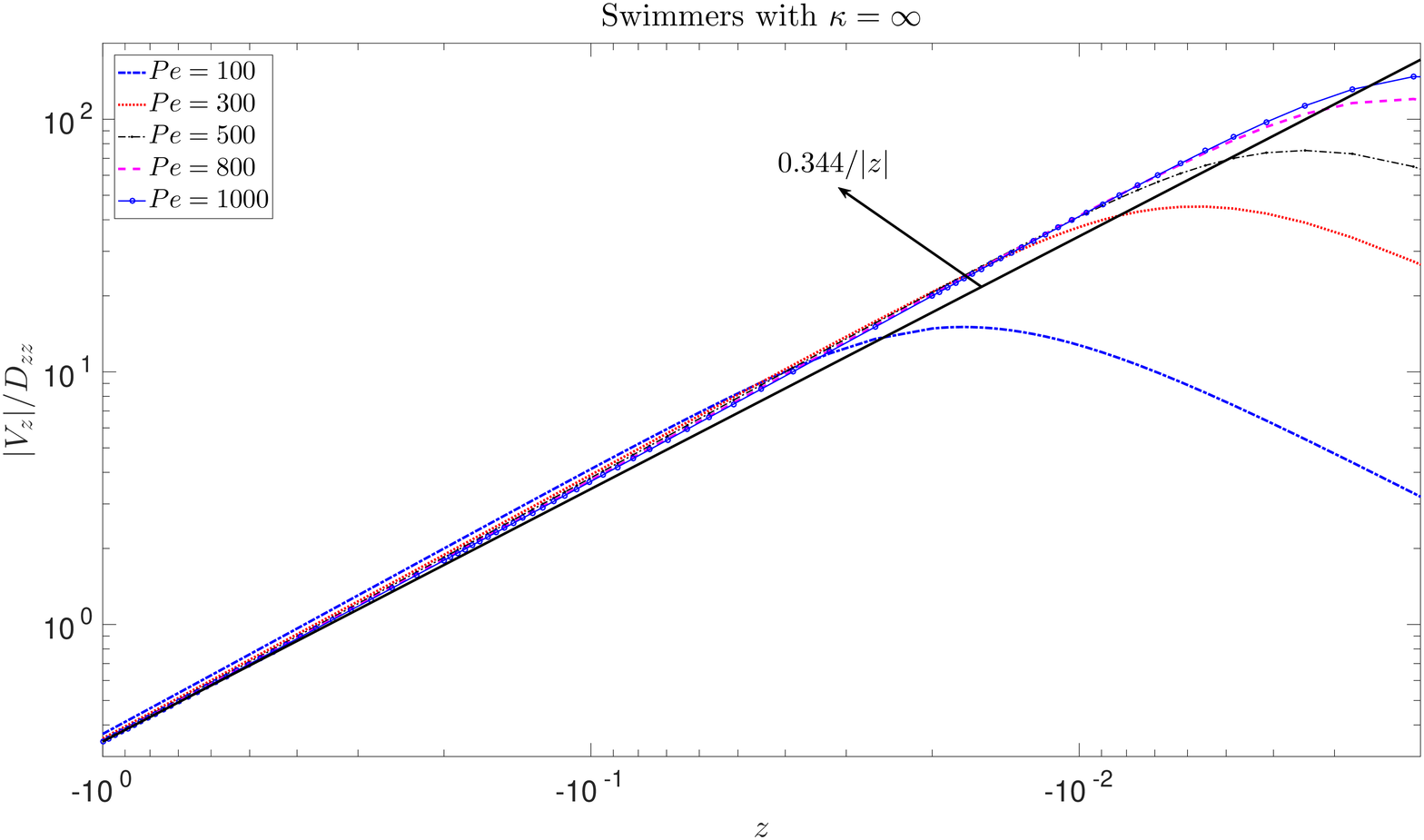}
\caption{Comparison of $\vert V_z\vert/D_{zz}$ against $z$, at different $Pe$'s for swimmers with $\kappa=\infty$. The black-solid line represents the large $Pe$ asymptotic profile of $\vert V_z\vert/D_{zz}$.}
\label{FIG:5}
\end{figure}

To further investigate the nature of the wallward migration, we plot, in figure \ref{FIG:11}, $D_{zz}$ and $V_z$ at the channel wall as a function of \textit{Pe} for $\kappa=\infty$ and 15. For $\kappa=\infty$, both $D_{zz}$ and $V_{z}$, for any non-zero $z$ scale as $Pe^{-4/3}$ for $Pe\gg 1$; being determined by $Pe_z$, the onset of this asymptotic scaling regime is postponed to a progressively larger $Pe$ with decreasing $z$ (not shown). For $\kappa=15$ too, there is an intermediate range of $Pe$'s ($\sim$ 3 to a little greater than 10) where the aforementioned scaling holds, but there is a deviation at larger $Pe$'s. As will be seen in \textsection\ref{Sec:3.2}, this deviation is a signature of the impending low-shear trapping regime. The observed scalings may be rationalized by noting that $D_{zz}\sim U^2_{z}t_{c}$, where $U_{sz}\sim $ O$ (U_{s}Pe^{-1/3})$ is the swimming speed, projected along the gradient direction, of a swimmer in the large-$Pe$ orientation boundary layer, with an angular extent of O($Pe^{-1/3}$), that forms around the flow direction [\cite{hinch1972effect, brenner1974}]; $t_c\sim \langle\theta^2\rangle/D_r\sim Pe^{-2/3}/D_r$ is the time over which rotary diffusion causes the swimmer orientation to fluctuate across the flow axis, leading to a de-correlation in its gradient-directed motion. The scaling for $V_z$ may be obtained by from arguments based on a gradient-directed biased random walk. One may write $V_z=\Delta z_{+} f_{+}-\Delta z_{-} f_{-}$, where $\Delta z_{+}$, $\Delta z_{-}$ represent the random walk jumps with frequencies $f_{+}$, $f_{-}$ along the positive and negative gradient directions, respectively. Expressing the random walk jumps as the ratio of swimming velocity in that direction to the associated frequency, one has $V_{z}=U_{sz}(z)-U_{sz}(z+\Delta z^{\prime})=U_{s}(Pe^{-1/3}_{z}-Pe^{-1/3}_{z+\Delta z^{\prime} })$, where $\Delta z^{\prime}$ is the (dimensionless) gradient-projection of the swimmer mean free path. With a swimming velocity of O($U_{s}Pe^{-1/3}_{z}$) and a decorrelation time of O($Pe^{-2/3}_{z}D^{-1}_{r}$), one obtains $\Delta z^{\prime} = U_{s}H^{-1}Pe^{-1}_{z}D^{-1}_{r}$. Using this leads to $V_z \sim$ O$(U^2_{s}Pe^{-4/3}_{z}H^{-1}D^{-1}_{r}/z) $.

The identical drift and diffusivity scalings above imply that the ratio $V_z/D_{zz}$ is independent of $Pe$ for large $Pe$, and hence, so is the concentration profile. We find
 $V_{z}/D_{zz}\propto 1/z$, consistent with the scaling arguments above, with a constant of proportionality ($C$) of order unity. This leads to a large-$Pe$ functional form of the concentration profile given by $\bar{F}=(1+C)\vert z \vert^{C}$. In figure \ref{FIG:5}, for $\kappa=\infty$, we compare the variation of $\vert V_z \vert/D_{zz}$ against $z$, for different $Pe$. A fit to the large-$Pe$ variation away from the centerline leads to  $C=0.344$.  From figure \ref{FIG:1}a, the swimmer concentration profiles are seen to approach the above $Pe$-independent limiting form for $Pe\geq 40$. Although, the steady state profile is $Pe$-independent, the time taken to attain steady state is of O($(H^2D_r/U^2_s)Pe^{4/3}$) and diverges for $Pe \rightarrow \infty$. The limiting concentration profile is singular with a cusp at the centerline ($z=0$). The cusp is, of course, an artefact of the $Pe=\infty$ limit. For any finite $Pe$, the multiple scales analysis breaks down in an asymptotically small interval of O($U_{s}/D_r$) in the vicinity of the centerline (keeping in mind our interpretation of $\epsilon$ as a swimmer Knudsen number, this layer in the vicinity of the centerline, where ballistic swimmer trajectories play an important role, may be termed a Knudsen layer in traditional kinetic theory parlance). Note that both the swimmer orientation distribution and concentration in this Knudsen layer may be influenced by the bound trajectories observed for deterministic swimmers observed in \cite{zottl2013periodic}.
    
\begin{figure}
\includegraphics[width=1\textwidth]{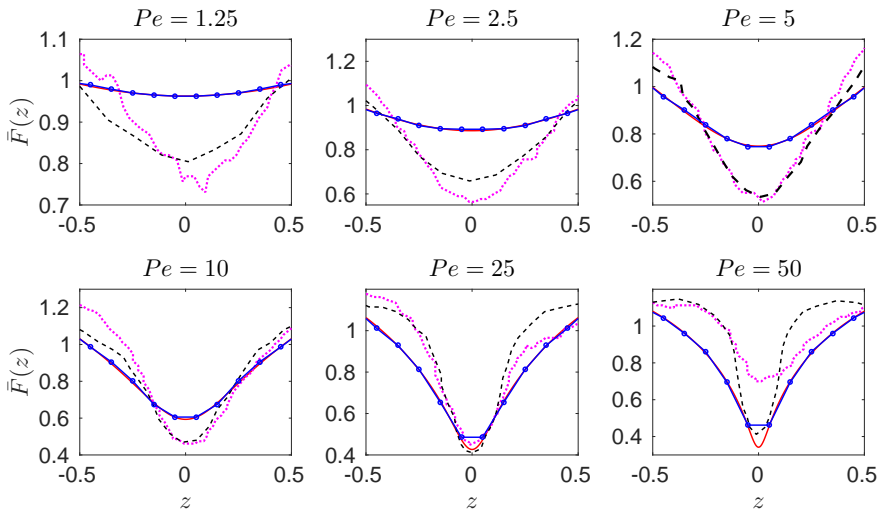}
\caption{Comparison of normalized concentration profiles: our theory (red line), Langevin simulation (blue line with circles) with the experiments (dotted line in magenta color) and simulations (black-dashed line) of \cite{rusconi2014bacterial}; $\kappa=10$.}
 \label{FIG:1Plus}
\end{figure}

In figure \ref{FIG:1Plus}, we compare our theoretical and simulation results (which are in mutual agreement regardless of $Pe$) with those reported in \cite{rusconi2014bacterial} for $\kappa=10$. For the three smallest $Pe$'s (1.25, 2.5 and 5), our theory under predicts the inhomogeneity compared to the experiments. There is improved agreement at $Pe=10$, and again, at the highest $Pe$'s (25 and 50), the disagreement with the experiments widens. The disagreement at the two highest $Pe$'s, we believe, is because the experiments [\cite{rusconi2014bacterial}] sampled under-developed concentration profiles. The channel length available for development of the concentration profile corresponds to a single arm of the serpentine tube used in the experiments, and is approximately 5 cm (note that, in light of the later experiments by Aronson and co-workers [\cite{sokolov2016rapid,sokolov2018instability}, in a curvilinear geometry, it is likely that the bends of the serpentine tube significantly disrupt the concentration profile developed in the straight sections). At large $Pe$, the entrance length may be estimated based on the time scale for attainment of steady state mentioned above, and is O($0.06\bar{u}H^2D_rPe^{4/3}/U^2_s$). Here, $\bar{u}=(2\bar{\dot{\gamma}}H/3)$ is the average suspension velocity and $\bar{\dot{\gamma}}=PeD_r$ is the the mean shear rate; the numerical pre-factor (0.06) in the above estimate is determined from the Langevin simulations. Based on this estimate, and using $H=212.5 \mu$m, $U_s=50 \mu$m $s^{-1}$, $D_r=1s^{-1}$ [\cite{rusconi2014bacterial}], the entrance length at $Pe = 50$ is 1.4m! For $Pe = 10$, the entrance length is about 3 cm, which is smaller than the aforementioned length of a straight section. This would suggest that the measured profile at $Pe = 10$ is indeed fully developed.
The fact that our simulation results at intermediate times closely match shapes of the experimental profiles at two largest $Pe$'s (not shown here), reinforces the above assertion of the latter profiles not being fully developed ones. A second reason for deviation from theory could be the role of wall interactions, which are not included in the present analysis, the focus here being on bulk mechanisms for  migration. Theory and simulations based on image induced interactions of pusher-type swimmers, and experiments on smooth swimming bacterial strains have shown swimmer accumulation at the boundaries in the absence of an imposed flow [\cite{berke2008hydrodynamic,lauga2009hydrodynamics, chilukuri2014impact}]. However, in the present context we expect this effect to only have a minor influence for two reasons. First, the time scale $t_{w}\sim$ O($L/U_s$), over which the wall interaction is significant, is comparable to the time scale for orientation decorrelation of O$(D^{-1}_{r})$ for wild-type swimmers under consideration, and the latter acts to disrupt any hydrodynamically induced accumulation for small $Pe$; in contrast, for the smooth swimmers used in \cite{berke2008hydrodynamic}, the decorrelation time scale given by O($D_{r-Brownian}^{-1}$), was greater by at least two orders of magnitude. At large $Pe$, rotary diffusion induced decorrelation is weak, and one then has a sustained image-induced interaction between nearly flow-aligned swimmers. Wall interactions, even in this case, are expected to only induce a concentration enhancement within a region of O($L$) from the walls. Using Brownian dynamics simulations for a confined channel geometry, with hydrodynamic interactions between the swimmers and wall included, \cite{chilukuri2014impact} showed that wall accumulation is indeed localized to the immediate vicinity of boundaries, consistent with the above arguments, and that this accumulation is further suppressed when the swimmers are subject to a pressure driven channel flow.
 Given that the observations of \cite{rusconi2014bacterial} and \cite{barry2015shear} were in the central portion of the channel, one expects the swimmer concentration to remain largely unaffected by wall interactions.

The disparity at the smallest $Pe$'s, pointed out above, could be due to multiple reasons. First is the possibility of the assumption $\epsilon \ll 1$, underlying the multiple scales analysis, breaking down. Based on the microchannel geometry and the known mean free path for \emph{B. subtilis} (estimated from the decorrelation time that is in turn inferred from the measured translational diffusivities), $\epsilon \approx 0.23$ in the experiments of \cite{rusconi2014bacterial}. Our simulations, which are not limited by the small $\epsilon$ assumption, agree well with theory even for $\epsilon = 0.2$, ruling out the lack of smallness of the swimmer Knudsen number as a cause of theory-experiment incongruity. Note that the issue of $\epsilon$ not being small enough is not relevant to the largest $Pe$'s since the stronger flow alignment of the swimmers with the unaltered decorrelation time implies that the projection of the mean free path in the cross-stream direction shrinks with increasing $Pe$, increasing the range of validity of the multiple scales analysis. For the high-shear trapping regime considered in this section, the effective Knudsen number is $\epsilon_{eff} = \epsilon Pe^{-1/3}$, the exponent reflecting the degree of swimmer alignment in the orientation boundary layer; for low-shear trapping examined next, $\epsilon_{eff} = \epsilon Pe^{-1}$ for large $Pe$. Next, it is also possible that the particular choice of the swimming mechanism determines the detailed concentration profile; for instance, there might be a significant difference between the degree of inhomogeneity in the swimmer concentration between RTP's and ABP's. As explained in the introduction, the former involves the swimmer undergoing a sudden large change in its orientation during tumbles [\cite{berg1993random, berg2008coli}], while the latter involves continuous small amplitude fluctuations in their orientation. Our Langevin simulations do suggest a greater inhomogeneity for RTP's than ABP's at the smallest $Pe$'s, with this difference diminishing in magnitude with increasing $Pe$. Thus, the difference between the swimming mechanism of \textit{B. subtilis}, relative to the pure ABP-dynamics assumed in the theoretical framework here, may cause the discrepancy between experiment and theory at the smallest $Pe$'s (the \cite{rusconi2014bacterial} experiments only measured the translation diffusivities, and do not therefore have detailed information with regard to the relative importance of tumbling vis-a-vis rotary diffusion). The sensitivity of the shear induced migration pattern to the swimming mechanism does deserve a more detailed study, and this will be reported in a future effort. 

\subsection{Swimmer migration towards the channel centerline - Low-shear trapping}\label{Sec:3.2}
\begin{figure}
\center
\includegraphics[width=1 \textwidth]{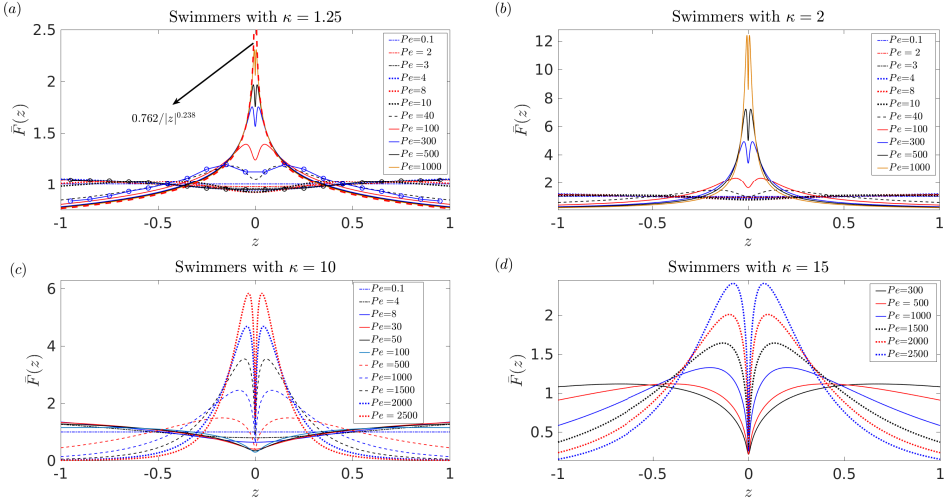}
\caption{Swimmer concentration profiles as a function of $Pe$ for different aspect ratios: (a) $\kappa=1.25$, (b) 2,  (c) 10, (d) 15. In (a), the red-dashed line is the large $Pe$ limiting profile, the solid-black and blue lines with circles represent the concentration profiles obtained from the Langevin simulations for $Pe=4$ and 40, respectively.}
\label{FIG:R1pt25ConcComp}
\end{figure}

Figure \ref{FIG:R1pt25ConcComp} shows the $Pe$-dependent swimmer concentration profiles for a range of swimmer aspect ratios, including the case $\kappa=15$ examined earlier (now over a larger range in $Pe$ extending upto 2500). For small to intermediate $Pe$, one observes high-shear trapping regardless of $\kappa$, consistent with the results of the small-$Pe$ analysis, and as seen before, the corresponding concentration profiles exhibit a single minimum at the centerline. At large $Pe$, however, there is a qualitative change with the concentration profiles now exhibiting a pair of maxima symmetrically disposed about the centerline; the maxima increase in amplitude, while converging towards the centerline with increasing $Pe$ (consistent with the earlier finite element computations of \cite{bearon2015trapping}). The aforementioned transition occurs across a $\kappa$-dependent threshold $Pe$. The Langevin simulations reinforce the existence of the transition in the concentration profiles; as shown in figure \ref{FIG:R1pt25ConcComp}a, the profile at $Pe=$4
exhibits a depletion of swimmers near the centerline, and the one at $Pe =$ 40 exhibits a near-center excess.
 
\begin{figure}
\center
\includegraphics[width=1 \textwidth]{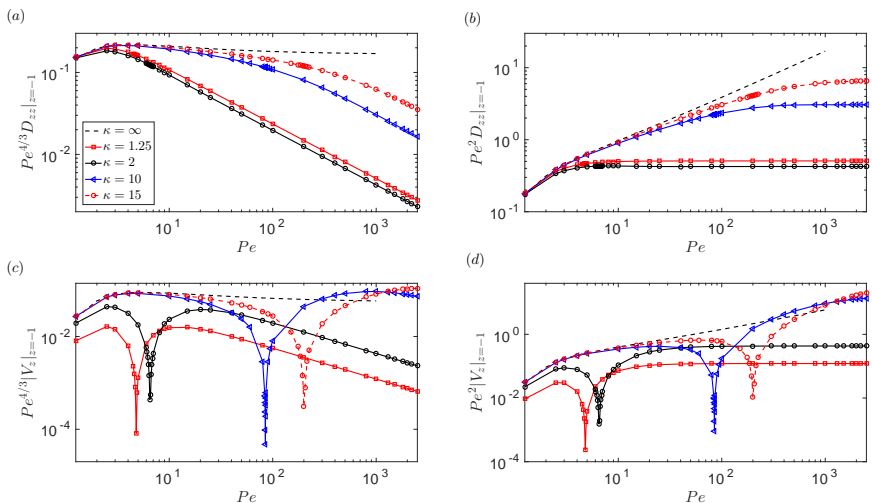}
\caption{The variation of (a)-(b) the diffusivity ($D_{zz}$) and (c)-(d) the absolute drift velocity ($\vert V_{z} \vert$) at the channel wall, as a function of $Pe$, for swimmers with different aspect ratios. Here, the $Pe^{4/3}$ and $Pe^2$ pre-factors, used to rescale the $D_{zz}$ and $V_{z}$, are motivated by the scalings in the high-shear ($Pe^{-4/3}$) and low-shear ($Pe^{-2}$) trapping regimes.} 
\label{FIG:DriftDiffScaling}
\end{figure}

\begin{figure}
\center
\includegraphics[width=1 \textwidth]{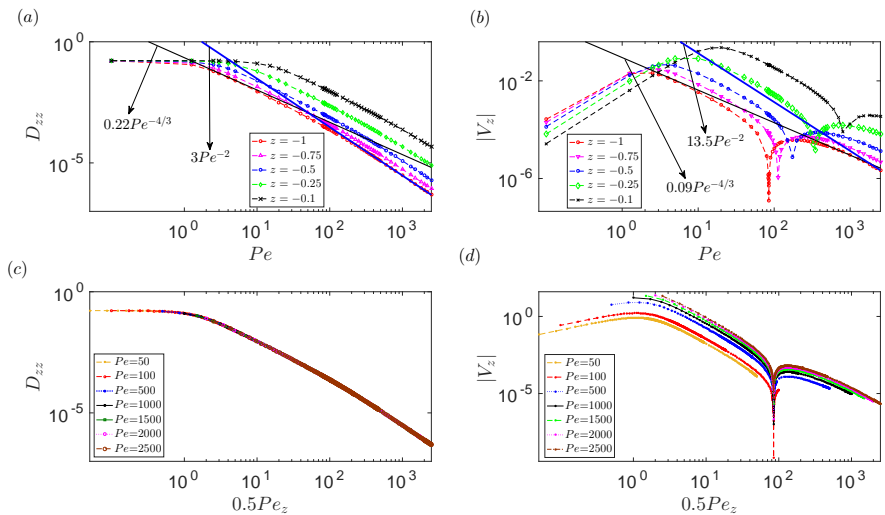}
\caption{The variation of (a) $D_{zz}$ and (b) $\vert V_z \vert$ against $Pe$ for swimmers with $\kappa=$ 10, at different locations across the channel. The black (thin) and blue (thick) lines in (a) and (b), represent the scalings of $D_{zz}$ and $\vert V_z \vert$ in the high-shear ($Pe^{-4/3}$) and low-shear ($Pe^{-2}$) trapping regimes, respectively. The variation of (c) $D_{zz}$ and (d) $\vert V_{z} \vert$ for $\kappa=10$ against $0.5Pe_z$, for different $Pe$; here, the factor 0.5 in front of $Pe_z$ is to ensure that $0.5Pe_z$ at the boundary is equal to $Pe$.}   
\label{FIG:R10DriftDiffScaling1}
\end{figure}

 To understand the transition from high-shear to low-shear trapping with increasing $Pe$, in figure \ref{FIG:DriftDiffScaling}, we plot $D_{zz}\vert_{z=-1}$ and $\vert V_z\vert_{z=-1}$, appropriately scaled,  against $Pe$ for $\kappa=\infty$ and for the $\kappa$'s in figure \ref{FIG:R1pt25ConcComp}. From figures \ref{FIG:DriftDiffScaling}c and d, $V_z$ is seen to undergo a change in sign with increasing $Pe$ (the associated zero-crossing appears as a sharp dip in the log-log plots), which is responsible for the reversal in migration. For $Pe$'s smaller than that corresponding to the zero-crossing (of $V_z$), swimmers migrate towards the walls; for larger $Pe$'s, swimmers migrate towards the centerline, leading to a low-shear trapping behavior. 
  It is important to note that the reversal in migration is actually a function of $Pe_z$, occurring for $Pe_z$ exceeding a $\kappa$-dependent threshold. This is confirmed in figure \ref{FIG:R10DriftDiffScaling1}, where the plots of the diffusivity and drift against $Pe$ for $\kappa = 10$, at different transverse locations ($z$), confirm the onset of the asymptotic O($Pe^{-2}$) scaling regime at progressively larger $Pe$ for $z$ approaching the centerline (figures \ref{FIG:R10DriftDiffScaling1}a and b). The reversal in drift is also delayed at locations closer to the channel centerline, with the earliest reversal occurring at the walls (the one shown in figure \ref{FIG:R10DriftDiffScaling1}b). To further confirm the role of $Pe_z$ (rather than $Pe$), in figures \ref{FIG:R10DriftDiffScaling1}c and d, $D_{zz}$ and $V_z$ are plotted as a function of $Pe_z$. This collapses the different $D_{zz}$ curves into a single one, and also collapses the zero-crossings of the different $V_z$ curves onto a single location corresponding to a critical $Pe_z$. Since $Pe_z$ always becomes arbitrarily small close enough to the centerline, there is always a region close to the centerline, regardless of $Pe$, where swimmers continue to exhibit a wallward migration. This leads to the non-monotonicity, with a pair of maxima bracketing a central dip in the concentration profiles shown in figure \ref{FIG:R1pt25ConcComp}; thus, the central portions of all the profiles in figure \ref{FIG:R1pt25ConcComp}, corresponding to low-shear trapping, still resemble the depletion profiles seen earlier is \textsection\ref{Sec:3.1}. It is only the limiting excess profile, for $Pe=\infty$, where swimmers migrate towards the centerline regardless of $z$ (this singular profile is subject to the limitations of the multiple scales analysis mentioned earlier in \textsection\ref{Sec:3.1}). As evident from figure \ref{FIG:DriftDiffScaling}, for any finite $\kappa$, both $D_{zz}$ and $V_z$ exhibit an O($Pe^{-2}$) scaling in the low-shear trapping regime that emerges for sufficiently large $Pe$, although for the larger values of $\kappa$ (10 and 15), there is the emergence of an intermediate scaling regime (evident from figures \ref{FIG:DriftDiffScaling}a and c) where $D_{zz}, V_z\sim Pe^{-4/3}$, already seen in \textsection\ref{Sec:3.1}. Thus, $\kappa=\infty$ is the only case where the O($Pe^{-4/3}$) scaling regime persists in the limit $Pe \rightarrow \infty$, as indicated by the black-dashed lines in figure \ref{FIG:DriftDiffScaling}. Infinitely slender swimmers therefore represent a singular limit with regard to shear induced migration behavior. Of course, for large enough $\kappa$'s, the $Pe$ corresponding to the transition to low-shear trapping may be large enough as to be numerically inaccessible. This is seen to be the case in figure \ref{FIG:DriftDiffScaling}b for $\kappa$=15, and in figure \ref{FIG:DriftDiffScaling}d for both $\kappa=10$ and $\kappa=15$; owing to the zero-crossing, it takes a larger $Pe$ for the drift to conform to the eventual O($Pe^{-2}$) scaling behavior.        
 
From the results presented thus far, it is seen that the transition from high to low-shear trapping occurs at a threshold $Pe$ of order unity for $\kappa$'s of order unity. For large $\kappa$, the cross-over $Pe$ is asymptotically large, and marks a transition from O($Pe^{-4/3}$) to an O($Pe^{-2}$) scaling regime for the drift and diffusivity coefficients (see figure \ref{FIG:R10DriftDiffScaling1}). From classical work on the orientation dynamics of passive anisotropic Brownian particles in a simple shear flow [ \cite{leal1971effect,hinch1972effect,brenner1974}], the former scaling regime implies the localization of the swimmer orientation distribution in an O($Pe^{-1/3}$) boundary layer around the flow direction where there is a balance between the rotary diffusion and flow-induced rotation. In contrast, when flow-induced rotation along Jeffery orbits occurs on a time scale much shorter than rotary diffusion, the latter acts as a regular perturbation; there is no orientational boundary layer, and one expects a scaling regime involving an integral powers of $Pe^{-1}$. Thus, the aforementioned cross-over $Pe$ must correspond to the point where the flow-induced (Jeffery) rotation, on a time scale of O($\dot{\gamma}^{-1}\kappa$) is comparable to that characterizing rotary diffusion of O($D^{-1}_{r}Pe^{-2/3}$); this leads to $Pe_c\sim$O($\kappa^3$) for $Pe \gg 1$ (this scaling is confirmed in \textsection\ref{subsec:PhaseDiag}, where we demarcate high and low-shear trapping on the $Pe-\kappa$ plane). Now, at leading order, one has purely flow induced Jeffery rotations of the swimmer, with an orbital time period $t_{jeff}\sim 1/(D_{r}Pe)$. The Jeffery rotations coupled with swimming lead to a swimmer trajectory, that along the cross-stream direction, has a bounded but oscillatory character with an amplitude $\Delta z$ of O($U_{s}/(D_{r}Pe)$). Rotary diffusion disrupts the exact periodicity of the leading order trajectory, causing an occasional slip of O($U_{s}/(D_{r}Pe)$). These random displacements in the $z$-direction, occurring over a time scale of O($D^{-1}_{r}$), leads to a $D_{zz} \sim \langle \Delta z^2\rangle/t_c$ which may be written as O($(U^{2}_{s}/D_{r})Pe^{-2}$), confirming the scaling behavior observed in the low-shear trapping regime; here, $\langle\Delta z^2\rangle$ is the mean square displacement of the swimmer in the gradient direction. Note that, for large $\kappa$, swimmers spend an O$(\kappa/(D_{r}Pe))$ time remaining nearly aligned with the flow axis at an angle of O($\kappa^{-1}$). This leads to the same estimate for $\Delta z$ as above, but a faster decorrelation rate of O($\kappa^{2}D_r$) (since rotary diffusion only has to induce an angular displacement of O($\kappa^{-1}$)), leading to a scaling for $D_{zz}$ of O($(U^{2}_{s}\kappa^2/D_{r})Pe^{-2}$). Based on the biased random walk argument as mentioned in \textsection\ref{Sec:3.1}, we can express $V_z=U_{sz(z)}-U_{sz(z+\Delta z^{\prime})}$. Now, for swimmers with $\kappa$ of O(1), using $U_{sz(z)}=U_{s}Pe^{-1}_{z}$, we obtain $V_{z}\sim$ O($(U^{2}_{s}H^{-1}D^{-1}_{r}/z)Pe^{-2}_{z}$). The drift and diffusivity scalings above are broadly consistent with those obtained in figure \ref{FIG:DriftDiffScaling}b.

\begin{figure}
\center
\includegraphics[width=1 \textwidth]{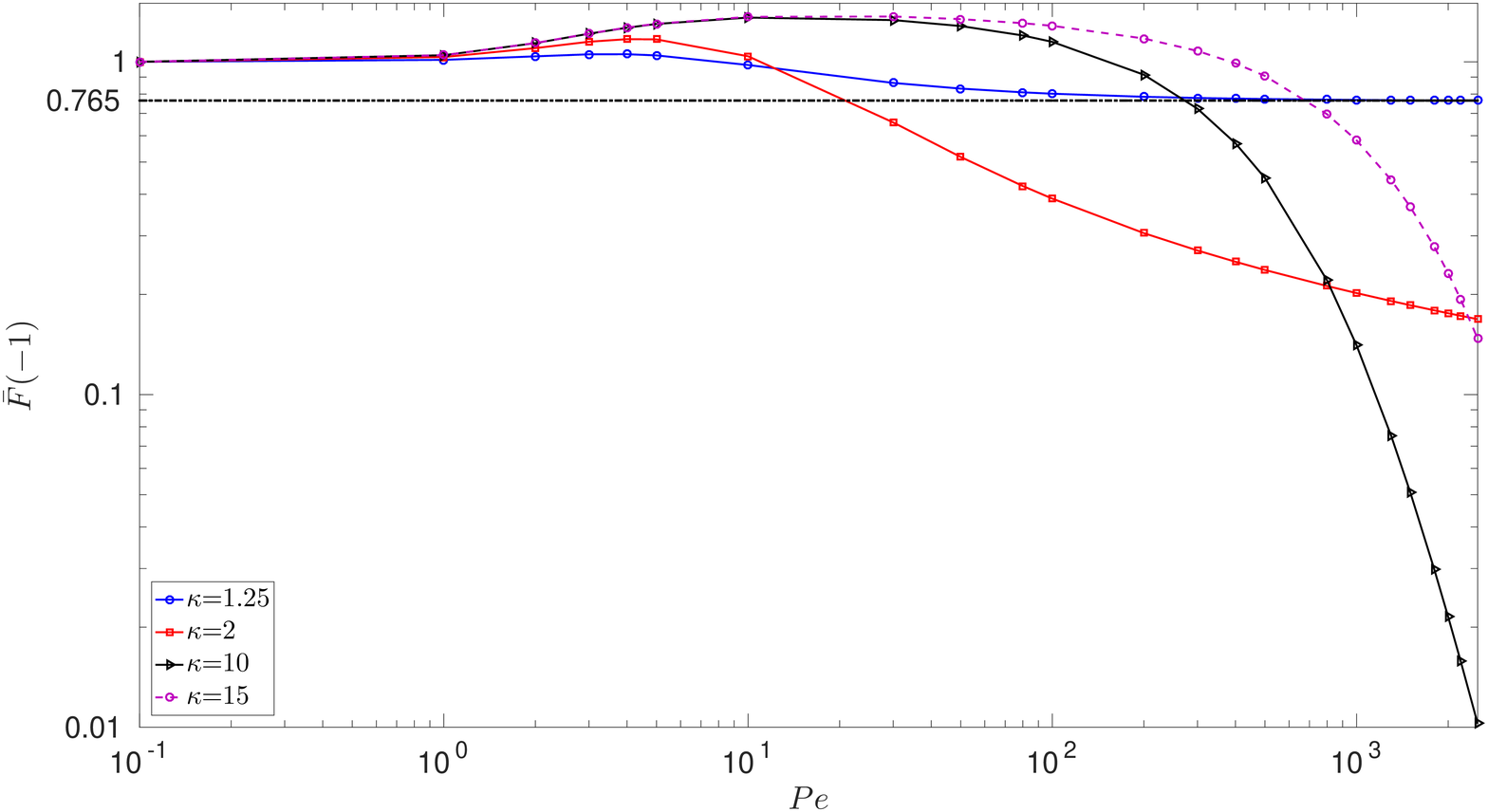}
\caption{Variation of swimmer concentration at the wall ($\bar{F}(-1)$) against $Pe$ for swimmers with different aspect ratios.}
\label{FIG:NearWallConcentration}
\end{figure}

For large $Pe$, similar to the high-shear trapping case, $V_z$ and $D_{zz}$ exhibit the same O($Pe^{-2}$) scaling, their ratio being independent of $Pe$ and of the form $C(\kappa)/z$, where $C$ is now a function of $\kappa$. One therefore expects a limiting $Pe$-independent profile of the form $(C+1)|z|^C$; note that, for any finite $Pe$, this profile breaks down in the vicinity of the centerline, where the actual finite-$Pe$ profiles exhibit a dip, corresponding to the onset of depletion (wallward migration) at small $Pe_z$. Since $C(\kappa)<0$ for low-shear trapping (owing to the sign-reversal of the drift), this limiting profile has a diverging concentration at the centerline. For $\kappa=1.25$, $C(\kappa)\sim-0.238$, and the associated limiting profile compares well to the numerically computed ones in figure \ref{FIG:R1pt25ConcComp}a. An interesting consequence of the change in sign of $C(\kappa)$, leading to low-shear trapping, is that the aforementioned limiting profile fails to be integrable for $C(\kappa)\leq -1$ (this wasn't an issue in the high-shear trapping case, since the swimmer concentration in all profiles, including the one corresponding to $Pe=\infty$, is bounded below by zero). We find $C(\kappa)$ to equal this threshold value (-1) for $\kappa\approx 2$, and decrease further for larger $\kappa$. The implication of the non-integrability above is that, for $\kappa \geq 2$, the swimmer concentration profiles must exhibit a `centerline collapse' in the limit $Pe \rightarrow \infty$; in other words, $\displaystyle{\lim_{Pe\rightarrow \infty,\kappa \geq 2}} \bar{F}(z)=\delta(z)$. Figure \ref{FIG:NearWallConcentration} shows the variation of swimmer concentration at the wall $(\bar{F}(-1))$ against $Pe$ for the aspect ratios considered in figure \ref{FIG:R1pt25ConcComp}. For $\kappa =$ 1.25, the near-wall swimmer concentration approaches a finite limiting value $\left(\displaystyle{\lim_{Pe\rightarrow \infty,\kappa=1.25}\bar{F}(-1)=0.765}\right)$. For $\kappa \geq 2$,  however, the near-wall swimmer concentration decreases monotonically to zero for $Pe \rightarrow \infty$, in accordance with the centerline-collapse hypothesized above. The slope characterizing this decrease appears to increase with increasing $\kappa$, implying that the collapse increases in intensity for larger $\kappa$ (for $\kappa =$ 15, the numerics have not accessed the asymptotic regime yet). The $\kappa$-dependence of the collapse does suggest that the drift and diffusivity coefficients must scale differently with $\kappa$ for $\kappa \gg 1$. Scaling arguments presented above suggested that $D_{zz} \sim $O$(\kappa^{2} Pe^{-2})$ in the low-shear trapping regime, and the large$-\kappa$ scaling of the drift must therefore differ from O($\kappa^2 Pe^{2}_{z}/z$) that would emerge from the biased-random-walk scaling above modified for $\kappa \gg 1$. The large-$\kappa$ scaling of the drift in this regime will be analyzed in detail in a future effort.

\begin{figure}
\center
\includegraphics[width=1 \textwidth]{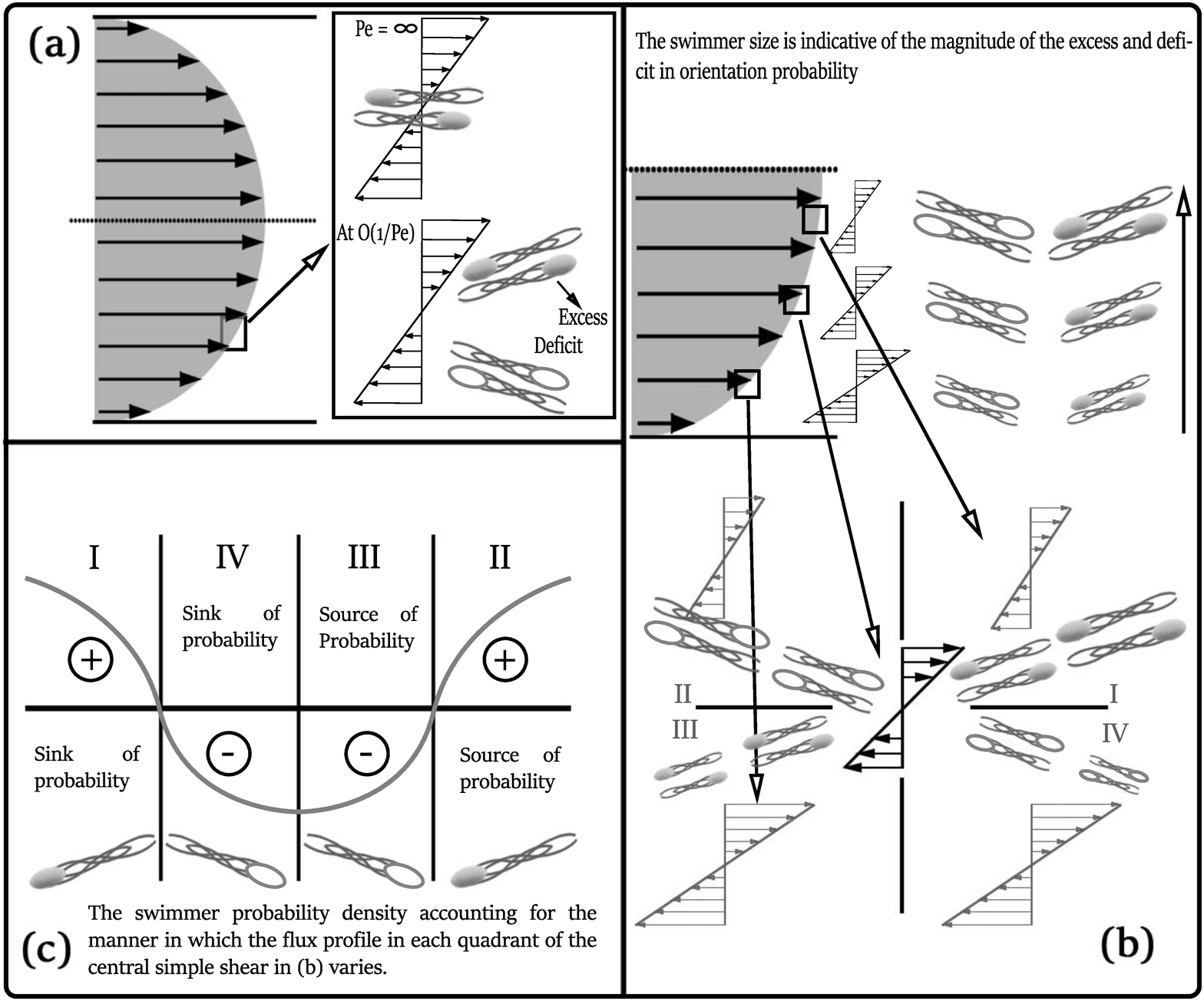}
\caption{A schematic highlighting the physical mechanism underlying the low-shear trapping regime. The solid-headed and hollow-headed swimmers represent the excess and deficit of swimmer (orientation) probability, respectively.}
\label{FIG:LowPeSchematic}
\end{figure}

Unlike the rather intuitive picture underlying high-shear trapping (see figure \ref{FIG:HighPeSchematic} in \textsection\ref{Sec:3.1}), the physical mechanism underlying low-shear trapping is more subtle. Here, we present arguments that support migration of swimmers towards the centerline at large $Pe$, beginning from the orientation dynamics of swimmers in a local linear flow description, as shown in figure \ref{FIG:LowPeSchematic}a. For the parabolic flow under consideration, the local linear approximation yields a simple shear flow with a shear rate that decreases from a maximum at the wall to zero at the centerline. For any finite $\kappa$, and sufficiently large $Pe$ ($Pe \gg 1$ for $\kappa$ of O(1); $Pe \gg \kappa^3$ for $\kappa \gg 1$), swimmers at any non-zero $z$ execute Jeffery rotations at leading order, with a time period commensurate with the $z$-dependent shear rate of the simple shear flow above. The resulting distribution of azimuthal angles within any Jeffery orbit is symmetrical about the flow direction (as first shown by \cite{leal1971effect}, the distribution across the orbits is still controlled by asymptotically weak rotary diffusion, although this doesn't affect the argument for the direction of drift given below). At the next order, O($1/Pe$), rotary diffusion breaks the aforementioned symmetry, leading to an excess of orientations in the extensional quadrants, and an equal and opposite deficit in the compressional ones (this antisymmetry is a consequence of Brownian motion being a weak but regular perturbation; in the singular case, relevant to the high-shear trapping regime, almost the entire swimmer probability is confined to the extensional quadrants). The aforementioned distributions at O(1) and O($1/Pe$) are sketched in figure \ref{FIG:LowPeSchematic}a. Owing to the decrease of $Pe_z$, the amplitudes of the excess and deficit above increase as one moves away from the wall. The lower figure in \ref{FIG:LowPeSchematic}b contains sketches of the local simple shear flows at three different $z$-locations, with the local excess and deficit in orientations in accordance with the arguments above (the swimmer lengths shown denote the amplitude of the excess or deficit). This picture may now be used to argue, in a quadrant-wise manner, for the swimming-induced changes the orientation distribution. At any $z$, in quadrant I, lesser excess swims in from the stronger shear flow below, and a greater excess exits into the weaker shear flow above; in quadrant III, the exact opposite occurs. Similarly, in quadrant II (IV), a greater (lesser) deficit exits into the weaker (stronger) shear flow above (below), with a lesser (greater) deficit swimming in from the stronger (weaker) shear flow below (above), leading to a net excess (deficit) of swimmers. As a result, there is a swimming-induced source of probability in quadrants II and III, and a sink in quadrants I and IV. In turn, this implies that the azimuthal flux of probability will, on a clockwise traversal, increase through quadrants II and III, and then decrease through IV and I. Accounting for the depleted orientation probability in the compressional quadrants (II and IV), this flux-profile leads to an excess of swimmers in quadrants I and II, implying a net drift towards low-shear regions. This migration towards the low-shear regions continues till the swimmers reach sufficiently close to the channel center where $Pe_z$ eventually falls below $\kappa^3$. From this location onward, and up until the centerline, the swimmers revert to a wallward migration. Therefore, the steady state concentration profile is such that it exhibits a maximum at a location intermediate between the centerline and the wall, with this location shifting towards the centerline with increasing $Pe$.

\subsection{Depletion index and the $Pe-\kappa$ migration portrait}\label{sec:depletion_index}
Having discussed the two kinds of migration patterns mentioned above, in \textsection\ref{subsec:DI}, we quantify the same in terms of the depletion index $DI$, a non-dimensional measure of the inhomogeneity of the swimmer concentration field: $DI<0~(>0)$ corresponds to high-shear (low-shear) trapping. Next, in \textsection\ref{subsec:PhaseDiag}, we organize the migration behavior on a $Pe-\kappa$ plane (for small $\epsilon$), thereby delineating the regimes of high-shear and low-shear trapping across a numerically determined phase boundary.  
 
\subsubsection{The depletion index (DI)}\label{subsec:DI}
\begin{figure}
\center
\includegraphics[width=1 \textwidth]{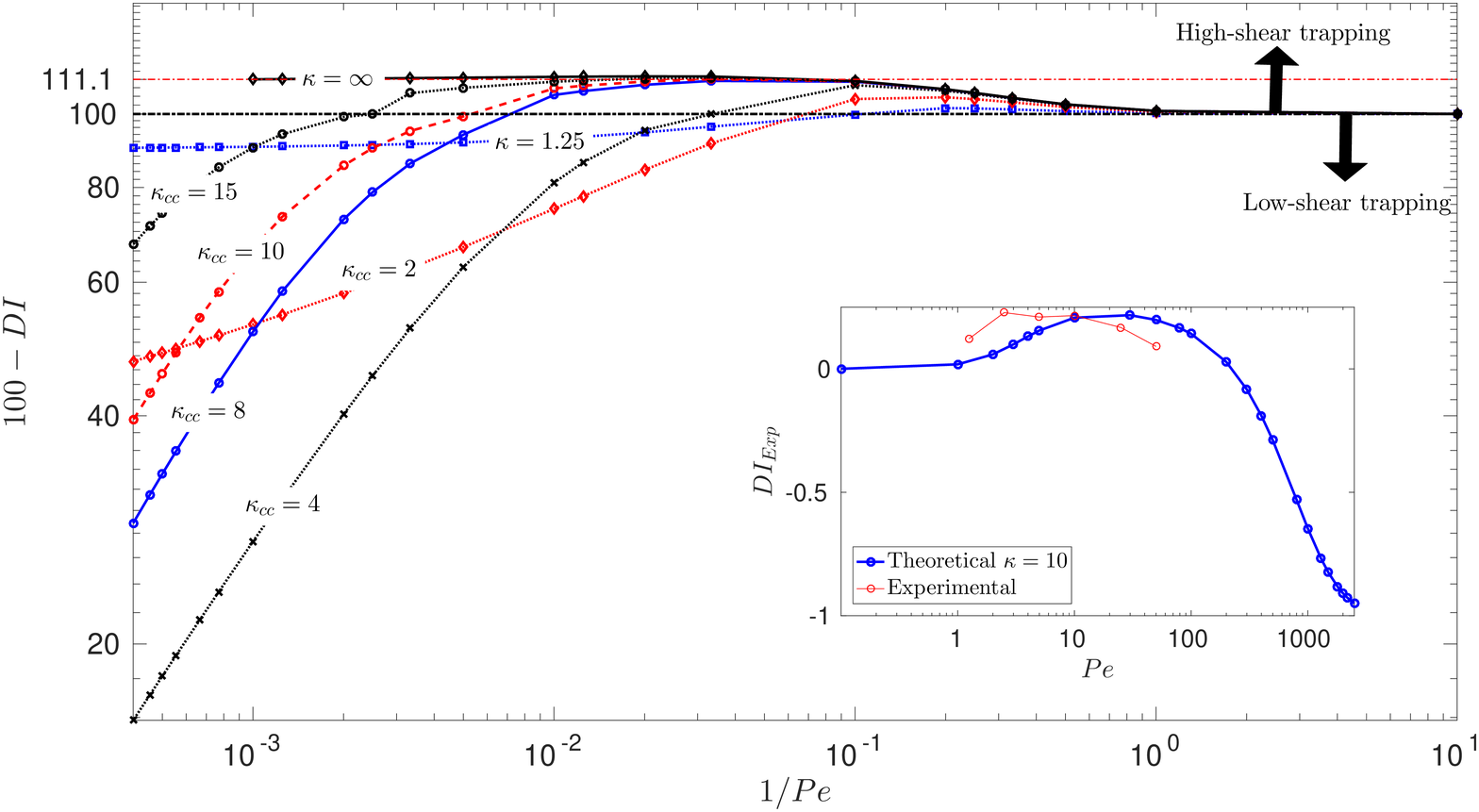}
\caption{Plot of $\Delta DI$ ($100-DI$) against $1/Pe$ for swimmers of different aspect ratios. The subscript `$cc$' in the aspect ratio ($\kappa_{cc}$) denotes  `centerline collapse'. The inset represents the comparison of $DI_{Exp}$ (defined separately; see text), as predicted by the analysis here, and the experimental results of \cite{rusconi2014bacterial}.}
\label{FIG:DeltaDI}
\end{figure}
We define $DI=2( A_1- A_2)/ A$, where $A_1$ represents the area under the initial uniform distribution from the wall up to its intersection with the steady state swimmer concentration profile ($z_{in}$), $A_2$ denotes the analogously defined area under the steady state profile from the channel wall to $z_{in}$ and $ A$ is the total area under either profile. For depletion profiles, conservation of swimmer number does imply that equality of areas between the wall and $z_{in}$, and between $z_{in}$ and centerline. To analyze swimmer migration at large $Pe$, $\Delta DI=100-DI$ is plotted against $1/Pe$ for swimmers of different aspect ratios in figure \ref{FIG:DeltaDI}. The plot discriminates between high ($\Delta DI>100$) and low-shear trapping behavior ($\Delta DI<100$), and in addition, helps differentiate between centerline-collapse ($\Delta DI \rightarrow 0$ for $1/Pe \rightarrow 0$ for $\kappa \geq 2$) and non-collapse ($\Delta DI$ approaches a finite value as for $1/Pe \rightarrow 0$ for $\kappa < 2$) behavior within the low-shear trapping regime. Figure \ref{FIG:DeltaDI} clearly illustrates the singular nature of infinitely slender swimmers. For $\kappa=\infty$, $\Delta DI$ plateaus at a finite value ($>100$) for $1/Pe \rightarrow 0$. In contrast, for any finite $\kappa \geq 2, \Delta DI$ approaches zero in the same limit, and the rate of approach, rather counter intuitively, increases with increasing $\kappa$. For large $\kappa$'s, this accelerated approach towards the origin begins after an intermediate extended plateau with $\Delta DI>100$ (the same plateau value as for $\kappa=\infty$). Note that \cite{rusconi2014bacterial}, likely motivated by experimental considerations, had defined depletion index ($DI_{Exp}$) based on the area between the actual and uniform profile in the central half width of the channel. The general character of figure \ref{FIG:DeltaDI} is, however, unchanged regardless of the definition.

It is worth reiterating the behavior of the depletion index, $\Delta DI$, as a function of the swimmer aspect ratio starting from infinitely slender swimmers. Regardless of $\kappa$, $\Delta DI$ starts from 100 for $Pe \ll 1$. For $\kappa=\infty$, $\Delta DI$ increases with increasing $Pe$, approaching a constant ($\Delta DI=111.1$) for $Pe \rightarrow \infty$. For large but finite $\kappa$, $\Delta DI$ again increases with increasing $Pe$, asymptoting to the aforementioned plateau value in the range $1\ll Pe \ll \kappa^3$,  and thereafter, decreases rapidly with further increase in $Pe$. This behavior continues with decreasing $\kappa$, with the $Pe$-interval corresponding to the intermediate plateau above progressively shrinking in extent, and the eventual rate of approach to zero at large $Pe$ also decreasing with decreasing $\kappa$. For the ABP's examined here, this trend continues until a critical aspect ratio of approximately 2, when $\Delta DI$ approaches zero only at a logarithmic rate for large $Pe$; that is, for $\kappa=2$, $\Delta DI$ is O($1/\ln{Pe}$) for $Pe \gg 1$. For $\kappa < 2$, $\Delta DI$ approaches a finite positive plateau (less than 100) for $Pe \rightarrow \infty$, with this plateau value approaching 100 as $\kappa$ approaches unity (spherical swimmers). It is worth noting that the infinite-aspect-ratio assumption is used routinely in the rheological context, when analyzing swimmers suspensions, and generally yields good agreement in between experiments and theory [\cite{Saintillan2010, saintillan2010extensional, lopez2015turning, nambiar2017stress, nambiar2018stress}]. In the present context, however, even for moderately large aspect ratios, $\kappa\sim 4-6$ (relevant in the biological context), at large enough $Pe$, the swimmers can, in fact, end up being trapped in the low-shear regions. Thus, the effective aspect ratio parameter appears to be of considerable importance in the context of shear-induced migration.

 Although our theoretical predictions and simulations are in mutual agreement, there is nevertheless a disagreement with the experiments of \cite{rusconi2014bacterial}; aspects of this disagreement were already elaborated on in \textsection\ref{Sec:3.1}. In order to illustrate the discrepancy, we have plotted $DI_{Exp}$, against $Pe$, as an inset in figure \ref{FIG:DeltaDI} for $\kappa = 10$. While the disagreement appears a matter of detail, in the sense that the $DI_{Exp}$ in the experiments peaks at a smaller $Pe$ ($\approx$ 10), and starts decreasing earlier, it is not! The interpretation in the original experiments was that the $DI_{Exp}$ would asymptote to zero for larger $Pe$. As discussed earlier, this decay of the $DI_{Exp}$ to zero, implying the approach of the swimmer concentration profile towards homogeneity, is a consequence of the experimental profiles not being fully developed on account of the limited residence time (ironically enough, the apparent peaking of the $DI_{Exp}$ at a finite shear rate in these experiments was motivation for subsequent experiments by Aronson and co-workers [\cite{sokolov2016rapid,sokolov2018instability}], in a curvilinear geometry, where the inhomogeneity in swimmer concentration was argued to persist at infinite $Pe$, albeit for entirely different reasons). As argued above, the $DI_{Exp}$ does not approach zero, for any aspect ratio, in the limit of large $Pe$. Thus, the apparent decay of the theoretical $DI_{Exp}$ is a signature of the transition to a low-shear trapping regime, and the theoretical $DI_{Exp}$ will eventually asymptote to -1, for the aspect ratio of 10 considered (which allows for a centerline collapse as $Pe \rightarrow \infty$).
   
\subsubsection{The $Pe-\kappa$ migration portrait}\label{subsec:PhaseDiag}
\begin{figure}
\center
\includegraphics[width=1 \textwidth]{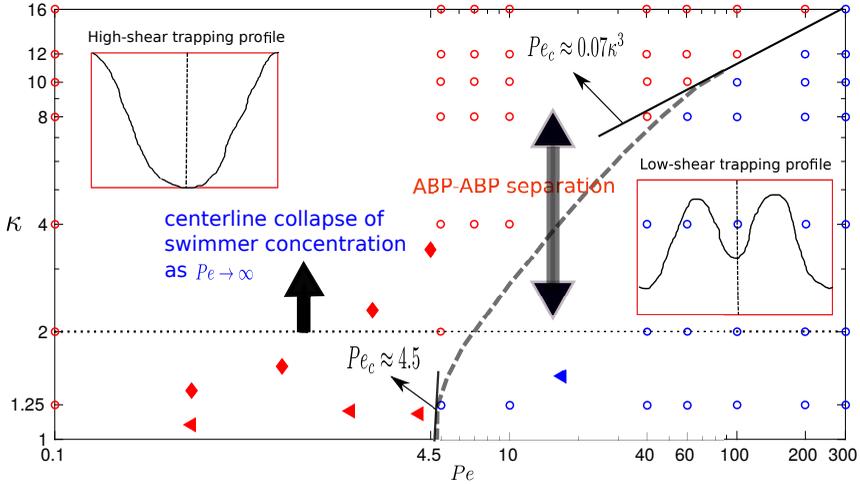}
\caption{Shear induced migration behavior on the $Pe-\kappa$ plane.
 The red and blue-circles, respectively, represent the high-shear trapping and low-shear trapping regimes (based on the profiles obtained from the multiple scales analysis, with partial verification from Langevin simulations), and the gray dashed line is an approximate boundary separating the two. The solid black lines denote the near-sphere and large-aspect-ratio asymptotes for this boundary. The diamonds and triangles respectively, corresponding to the different trapping regimes of algae \textit{Heterosigma} and \textit{Dunaliella} observed by \cite{barry2015shear}.} 
\label{FIG:PhaseDiag}
\end{figure}

Figure \ref{FIG:PhaseDiag} organizes the high-shear and low-shear trapping regimes on the $Pe-\kappa$ plane, with an approximate boundary separating the two. We differentiate the high-shear trapping from the low-shear trapping, based on the slope of the swimmer concentration profile at the channel wall. Across the phase boundary, there is a change in sign of this slope, arising from the reversal in the drift; see figure \ref{FIG:DriftDiffScaling} in \textsection\ref{Sec:3.2} (note that this criterion suffices in the present context where the focus is on shear induced migration in the bulk; inclusion of wall interactions, would require a modification of this criterion. This is not an issue for comparison with the experiments, however, since the swimmer concentrations were monitored well away from the boundaries). The large aspect ratio asymptote for the aforementioned boundary between the two trapping regimes has already been determined based on the scaling arguments given in \textsection\ref{Sec:3.2}, and is of the form $Pe_c \propto \kappa^{3}$; the pre-factor 0.07 given in figure \ref{FIG:PhaseDiag} is determined from a best fit for the curve separating the numerically determined depletion and excess profiles. Now, due to the absence of any preferential alignment under shear, spherical-swimmer concentration profiles are homogeneous regardless of $Pe$, in the dilute non-interacting regime under consideration. Nevertheless, the small but finite inhomogeneity that arises in the limit $\kappa \rightarrow 1$ can transform from a near-center depletion to a near-center excess at a finite $Pe$. This is indeed the case, and based on an expansion of the swimmer probability density in the small Bretherton constant $B$, as $\Omega=F/4\pi+ B \Omega_{1}+\cdots$, one finds $Pe_c \sim 4.5$ (see appendix \ref{App:Near-Sphere}). Therefore, as $\kappa\to 1$, ABP's exhibit high and low-shear trapping profiles across $Pe_c = 4.5$, with the analytical forms for the latter profiles also exhibiting the characteristic pair of maxima on either side of the centerline, already seen in the numerical profiles. This transition for near-spherical swimmers has again been validated using numerical predictions based on the multiple scales analysis, and Langevin simulations.

Also shown in figure \ref{FIG:PhaseDiag} are the results from the experiments of \cite{barry2015shear} corresponding to both high-shear and low-shear trapping of the phytoplankton algae \textit{Heterosigma} and \textit{Dunaliella}. The effective swimmer aspect ratios ($\kappa$) in the experiments were determined from a best fit of the theoretical orientation distribution, obtained by solving the Fokker-Planck equation with $\kappa$ as a fitting parameter, with the experimentally determined one. Note that the $\kappa$'s determined in the manner above were functions of $Pe$, implying that the swimmer shape changed on account of the shear induced deformation. Based on the concentration profiles and depletion index plots ($DI_{Exp}$) reported in \cite{barry2015shear}, for $Pe \leq 5$, \textit{Heterosigma} exhibit high-shear trapping, whereas, \textit{Dunaliella}, show high-shear trapping till $Pe\sim 5$ but transition to low-shear trapping at at $Pe\sim 12.5$. Similar to the observation of \cite{barry2015shear}, for the algae \textit{Dunaliella}, we too observed the transition from the high-shear to low-shear trapping with increase in $Pe$. The experimental data points are thus consistent with the trapping regimes identified in figure \ref{FIG:PhaseDiag}, and this consistency serves as an important validation of the analysis and arguments presented here.

\section{Conclusions}\label{Conclusion}
Motivated by recent microfluidic experiments [\cite{rusconi2014bacterial,barry2015shear}], in this work we have used both theory and simulations to systematically study shear induced migration in a dilute suspension of microswimmers subject to a plane Poiseuille flow in a wide channel. On the theoretical front, we have used the method of multiple time scales to derive a drift-diffusion equation governing the steady state cross-stream swimmer concentration profile. Based on the swimmer geometry, as characterized by its equivalent aspect ratio $\kappa$, and a rotary P\'eclet number $Pe = U_f/(D_r H)$ that characterizes the relative  efficiency of stochastic reorientation and flow-induced rotation, we delineate high-shear and low-shear trapping regimes. Our theoretical predictions are reinforced by the results of  Langevin simulations, and establish for the first time, the existence of a low-shear trapping regime at sufficient large $Pe$ for any finite $\kappa$. The transition from high-shear to low-shear trapping appears consistent with the experimental data available from \cite{barry2015shear}; the biflagellated algal species, \textit{Chlamydomonas} and \textit{Dunaliella}, used in the experiments exhibit a low-shear trapping regime at high shear rates, with the latter species going through a high-shear trapping regime at lower shear rates. The approach of microswimmers to any surface is the first step towards colonization and biofilm formation. The existence of both high- and low-shear trapping regimes points to the possibility of manipulating fluid shear to possibly reinforce or retard biofilm formation; thus, the high and low-shear trapping regimes in figure \ref{FIG:PhaseDiag} may perhaps be equivalently interpreted as regimes where hydrodynamics favors and opposes biofilm formation. A richer range of concentration patterns and dynamical regimes, relative to those identified earlier in \cite{rusconi2014bacterial} and \cite{bearon2015trapping}, is expected in the presence of an ambient chemical gradient that causes chemotactic swimmers to drift towards a boundary (as a possible precursor for biofilm formation), while a high-$Pe$ shear drives migration in the opposite direction. This particular aspect will be pursued in the near future.

 The organization of shear induced migration of active swimmers as a $Pe-\kappa$ phase portrait has important implications. The fact that swimmers on either side of the phase boundary in figure \ref{FIG:PhaseDiag} migrate in opposite directions allows, in principle, for a microfluidic shape sorter ($\kappa$ being a surrogate for the swimmer shape). A shape-based sorting might be feasible in the low-shear trapping regime alone by exploiting the centerline collapse (for $Pe\rightarrow \infty$) that occurs for $\kappa\geq 2$, (application of the multiple scales analysis shows that results for the plane Poiseuille flow derived herein, including the threshold aspect ratio for centerline collapse, directly carry over to the pipe Poiseuille case). A second interesting implication of the centerline collapse is for the dispersion of microswimmers in the flow direction (this has been examined, in the limit of confined channels, by \cite{chilukuri2015dispersion}). It is known from classical Taylor-dispersion theory that, in the limit where the microswimmers uniformly sample the channel cross-section, the mean square displacement in the flow direction scales as O($Pe_{t}^{2}D_{zz}t$), and therefore, increases with increasing $Pe_t$; here, $Pe_t = U_{f}H/D_{zz} = Pe^3/\epsilon^2$, is the P\'eclet number based on the swimmer translational diffusivity. For large-aspect-ratio microswimmers in the low-shear trapping regime identified here, the extent of the channel cross-section sampled decreases sharply with increasing $Pe_t$ (or increasing $Pe$ with $\epsilon$ fixed), pointing to a possible non-trivial scaling of the flow-induced dispersion in the limit of large $Pe_t$.
 
  An interesting consequence of the detailed analysis here is the remarkable sensitivity of the shear induced migration to subtle changes in the orientation dynamics. These differences in orientation dynamics were originally identified in the context of passive anisotropic particles, in a series of studies by Hinch and Leal [\cite{leal1971effect, hinch1972effect, leal1972rheology, hinch1972note, leal1973theoretical, hinch1973time}], who were motivated by the need to understand the subtle role of weak Brownian motion in strong shearing flows, and its implications for suspension rheology. In fact, they turn out to much more crucial for active particles, being responsible for the onset of a low-shear trapping regime! Indeed, \cite{nitsche1997shear} have previously used the method of multiple scales to examine shear induced migration in a dilute suspension of passive anisotropic particles. Such particles do exhibit a high-shear trapping on account of the anisotropy of the mobility, and thence, the Brownian diffusivity. The trapping arises because of the lower transverse diffusivity(mobility) of flow-aligned particles close to the walls, but it is a relatively weak effect owing to the mobility anisotropy (defined as the ratio of the longitudinal to the transverse mobility coefficients) saturating at 2 even for infinitely slender particles; experiments on fiber suspensions have failed to observe the predicted high-shear trapping [\cite{nitsche1997shear}] at lower volume fractions [\cite{strednak2018shear}]. Since the Brownian mobility, and therefore, the diffusivity  at any non-zero $z$, asymptotes to a finite value for $Pe \rightarrow \infty$, the concentration profile must become spatially homogeneous in this limit; in sharp contrast to the active case examined here.
 
  Assuming an independent source of translational diffusion (thermal or otherwise), with components $D^t_{\parallel}$ and $D^t_{\perp}$ along and transverse to the flow direction, in addition to the translational diffusion arising from swimming activity considered here, one may use scaling arguments to predict the transition from the active shear induced migration analyzed here to the migration behavior of passive anisotropic particles analyzed by \cite{nitsche1997shear}. This transition is most relevant to large aspect ratio swimmers, since swimmers with moderate $\kappa$ would have transitioned to their asymptotic low-shear trapping regime (collapse or non-collapse) at fairly moderate values of $Pe$. To estimate the threshold for the active-passive transition, we balance $D^t_{\perp}$ with the high-$Pe$ asymptote for the translational diffusivity in the high-shear trapping regime, given by $\sim 0.22 Pe^{-4/3}U^{2}_{s}/D_r$ (see figure \ref{FIG:R10DriftDiffScaling1}); we consider the diffusivity at the wall, which gives an upper bound for the threshold. The balance yields $Pe_c \sim [0.22 U_s^2/(D_r\,D^t_{\perp})]^{3/4}$. Equating this threshold to $\kappa^3$, one obtains a threshold aspect ratio $\kappa_c \sim [0.22 U_s^2/(D_r\,D^t_{\perp})]^{1/4}$  (as must be the case, the same expressions for $Pe_c$ and $\kappa_c$ result on using $D_{zz} \sim Pe^{-2}\kappa^2 U^{2}_{s}/D_{r}$ corresponding to the low-shear trapping regime). The significance of $\kappa_c$ above is that the concentration profiles for swimmers with $\kappa > \kappa_c$ would transition to homogeneity (for $Pe \rightarrow \infty$) directly from the high-shear trapping regime, while the concentration profiles for swimmers with $\kappa < \kappa_c$ will asymptote to homogeneity from the low-shear trapping regime. When $D^t_{\perp}$ has a thermal origin, using the familiar Stokes-Einstein expression with the translational mobility for a slender body, one has $D^t_{\perp} = k_b T\ln\kappa/(4\pi\eta L)$ [\cite{dhont1996introduction}] with $k_b$ being the Boltzmann constant, $T$ being the temperature, and $\eta$ being the suspension viscosity. This yields $\kappa_c\sim$O(10) for $L\approx9 \mu$m, $U_s\approx 50 \mu$m, $\kappa=10$, $T\approx 300$ and $\eta \approx 10^{-3}$ for the aqueous medium used to suspend the microorganisms [\cite{rusconi2014bacterial, lopez2015turning}]. Thus, the \textit{B. subtilis} in \cite{rusconi2014bacterial} experiments are expected to transition to a passive-migration behavior at a $Pe$ that is about two orders of magnitude higher than the highest $Pe$ sampled in the experiments. The larger algae used in \cite{barry2015shear} should yield much larger $\kappa_c$ values.

Finally, it must be noted that figure \ref{FIG:PhaseDiag} depicts migration behavior specifically for ABP's. One expects a similar phase portrait for other swimmer types (RTP's, for instance), although the aforementioned sensitivity to the orientation dynamics obviously points to different $Pe-\kappa$ scalings for the phase boundary (and a different threshold aspect ratio for a possible centerline collapse). Such an analysis, and its implications towards flow-based separation of microswimmers, with different swimming characteristics, will be reported separately.
\section*{Acknowledgements}
L.N.Rao would like to thank Science and Engineering Research Board, India (Grant No. PDF/2017/002050) for the financial support.
\begin{appendix}
\section{Numerical scheme for the swimmer probability densities and concentration profile.}{\label{App:FinitePe}}

In the following, we describe the numerical scheme adopted to evaluate the swimmer orientation probability densities $G(\mathbf{p})$ and $\Omega_1(\mathbf{p})$, at O(1) and O($\epsilon$), respectively, and thence  the steady state swimmer concentration profile of (\ref{EQ:DriftDiffEquation}), mentioned in \textsection\ref{subsec:conc}. To solve (\ref{EQ:Omega0}), for $G(\mathbf{p})$, we use a spherical coordinate system with its polar axis aligned with the $z$-axis (see figure \ref{FIG:Schematic}). As a result, the components of the swimmer orientation vector are given by $p_1=\sin \theta \cos \phi$, $p_2=\sin \theta \sin \phi$, $p_3=\cos \theta$. The flow-induced rotation term in (\ref{EQ:Omega0}), $\nabla_{p}\cdot(\tilde{\dot{\mathbf{p}}}G)$, may now be written as:
 
 \begin{eqnarray}
 \nabla_{p}\cdot(\tilde{\dot{\mathbf{p}}}G)&=&-3 B \sin \theta \cos \theta \cos \phi G+\frac{1}{2}\left(1+B \cos(2\theta) \right)\cos \phi \frac{\partial G}{\partial \theta}-\frac{1}{2} \left(1+B \right)\cot \theta \sin \phi \frac{\partial G}{\partial \phi} \nonumber \\ 
 &=& -3 B \sin \theta \cos \theta \cos \phi G+\frac{1}{2}\left[i \mathcal{L}_{y}(G)\lbrace 1+B \cos (2\theta) \rbrace- 2 i B \sin \theta \cos \theta \sin \phi \mathcal{L}_{z}(G) \right], \nonumber\\
\label{EQ:GExpansion}
  \end{eqnarray}
where the operators [\cite{messiah1962quantum,doi1978dynamics}]
\begin{eqnarray}
\mathcal{L}_{y}&=&-i \cos \phi \frac{\partial}{\partial \theta}+i \cot \theta \sin \phi \frac{\partial}{\partial \phi}, \nonumber \\
\mathcal{L}_{z}&=& -i \frac{\partial}{\partial \phi}.
\label{EQ:CleschOpearators}
\end{eqnarray}

Expressing (\ref{EQ:GExpansion}) in terms of spherical harmonics, we have:   
\begin{eqnarray}
 \nabla_{p}\cdot(\tilde{\dot{\mathbf{p}}}G) = -3 \kappa \sqrt{\frac{2\pi}{15}}(Y_{2}^{-1}-Y_{2}^{1}) G &+&\frac{1}{2}\left[i \mathcal{L}_{y}(G)\left\{ 1+\kappa \left(-\frac{2}{3}\sqrt{\pi}Y_{0}^{0}+\frac{8}{3}\sqrt{\frac{\pi}{5}}Y_{2}^{0}\right) \right\} \right. \nonumber \\
  &+& \left. \kappa 2  \sqrt{\frac{2\pi}{15}}\left(Y_{2}^{-1}+Y_{2}^{1} \right) \mathcal{L}_{z}(G) \right].
\label{EQ:Clbesch1}
\end{eqnarray}
Now, expanding 
\begin{equation}
G(\mathbf{p})=\sum_{l=0}^{\infty}\sum_{m=-l}^{l}a_{l,m}Y_{l}^{m}(\mathbf{p})
\label{EQ:GpExp}
\end{equation}
and substituting in (\ref{EQ:Clbesch1}), one may write:
\begin{align}
\nabla_{p}\cdot(\tilde{\dot{\mathbf{p}}}G)=\sum_{l=0}^{\infty}\sum_{m=-l}^{l}a_{l,m} \nabla_{p}\cdot(\tilde{\dot{\mathbf{p}}}Y_{l}^{m}) =  \sum_{l=0}^{\infty}\sum_{m=-l}^{l}a_{l,m} \left[ -3 B \sqrt{\frac{2\pi}{15}}(Y_{2}^{-1}-Y_{2}^{1}) Y_{l}^{m} \right. \nonumber \\
+\frac{i}{2}  \mathcal{L}_{y} \vert Y_{l}^{m}> \left\{ 1+B \left(-\frac{2}{3}\sqrt{\pi}Y_{0}^{0}+\frac{8}{3}\sqrt{\frac{\pi}{5}}Y_{2}^{0}\right)  \right\} 
  +\left. B   \sqrt{\frac{2\pi}{15}}\left(Y_{2}^{-1}+Y_{2}^{1} \right) \mathcal{L}_{z}\vert Y_{l}^{m}> \right],
\label{EQ:Clbesch2}
\end{align}
where $Y_{l}^{m}(\mathbf{p})=\sqrt{(2l+1)/(4\pi)(l-m)!/(l+m)!} P_{n}^{m}(\cos \theta) \exp{(im\phi)}$, represent the spherical harmonics and $P_{n}^{m}$ are the associated Legendre functions [\cite{abramowitz1965handbook}] and $\mathcal{L}_{i} \vert Y_{l}^{m}>=\mathcal{L}_{i} (Y_{l}^{m}) $.

Next, we substitute the expansion for $G(\mathbf{p})$ above in the remaining terms of (\ref{EQ:Omega0}), with the tumbling kernel being given by $\displaystyle K(\mathbf{p}/\mathbf{p}^{\prime})=\beta \exp{(\beta(\mathbf{p}\cdot \mathbf{p}^{\prime}))}/(4\pi \sinh \beta)=\sum_{n=0}^{\infty}A_{n}P_{n}(\mathbf{p}\cdot\mathbf{p}^{\prime})$ [\cite{subramanian2009critical,nambiar2017stress}]; here, $\beta$ is the correlation parameter characterizing the tumbles, and $P_n$ represent the Legendre polynomial of the $n$ th degree, with argument being the cosine of the angles between the pre- and post-tumble operations. Further, employing the addition theorem of spherical harmonics and that $\nabla^{2}_{p}Y_{l}^{m}=-l(l+1)Y_{l}^{m}$ [\cite{abramowitz1965handbook}], (\ref{EQ:Omega0}) may finally be written as the following summation over the spherical harmonics:

\begin{align}
 \sum_{l=0}^{\infty}\sum_{m=-l}^{l}a_{l,m} &\left[ -3 Pe_{z}B \sqrt{\frac{2\pi}{15}}(Y_{2}^{-1}-Y_{2}^{1}) Y_{l}^{m} + Pe_{z}\frac{i}{2} \mathcal{L}_{y} \vert Y_{l}^{m}> \left\{ 1+B \left(-\frac{2}{3}\sqrt{\pi}Y_{0}^{0}+\frac{8}{3}\sqrt{\frac{\pi}{5}}Y_{2}^{0}\right)  \right\} \nonumber \right. \\
  &\left.  +B Pe_{z}   \sqrt{\frac{2\pi}{15}}\left(Y_{2}^{-1}+Y_{2}^{1} \right) \mathcal{L}_{z}\vert Y_{l}^{m}> +l(l+1)Y_{l}^{m} \right.\\
    &\left. +\frac{1}{\tau D_r}\left(Y_{l}^{m}-\sum_{s=0}^{\infty}\sum_{t=-s}^{s}\int d \mathbf{p} ^{\prime} A_{s}\frac{4\pi(-1)^{t}Y_{s}^{t}(\mathbf{p})Y_{s}^{-t}(\mathbf{p}^{\prime}) Y_{l}^{m}(\mathbf{p}^{\prime})  }{(2s+1)}   \right)  \right]=0.
\label{EQ:G0Eq1}
\end{align}

In the above equation, $A_0=1/4\pi$ (from the conservation of swimmers during tumble events: $\int d \mathbf{p} K(\mathbf{p}/\mathbf{p}^{\prime})=\int d \mathbf{p}^{\prime} K(\mathbf{p}/\mathbf{p}^{\prime})=1 $) and 
\begin{equation}
A_n=\frac{\int d (\mathbf{p}\cdot \mathbf{p}^{\prime}) \beta \exp{(\beta(\mathbf{p}\cdot \mathbf{p}^{\prime}))}/(4\pi \sinh \beta) P_{n}(\mathbf{p}\cdot \mathbf{p}^{\prime})}{\int d (\mathbf{p}\cdot \mathbf{p}^{\prime}) P_{n}(\mathbf{p}\cdot \mathbf{p}^{\prime}) P_{n}(\mathbf{p}\cdot \mathbf{p}^{\prime})}~~~ \text{for}~~~n\geq 1. 
\end{equation}
Using the following identities in (\ref{EQ:G0Eq1}) [\cite{doi1978dynamics,messiah1962quantum}] 
\begin{eqnarray*}
i\mathcal{L}_{y}\vert Y_{l}^{m}>&=&\frac{1}{2}\left[k^{\prime}_{lm}Y_{l}^{m+1}-k^{\prime \prime}_{lm}Y_{l}^{m-1}\right], \nonumber \\
\mathcal{L}_{z}\vert Y_{l}^{m}>&=& mY_{l}^{m},
\end{eqnarray*}
and after simplification, we get:
\begin{align}
 \sum_{l=0}^{\infty}\sum_{m=-l}^{l}a_{l,m} &\left[ -3 B Pe_{z} \sqrt{\frac{2\pi}{15}}(Y_{2}^{-1}-Y_{2}^{1}) Y_{l}^{m} +\frac{ Pe_{z}(k^{\prime}_{lm}Y_{l}^{m+1}-k^{\prime\prime}_{lm}Y_{l}^{m-1})}{4}  \right. \nonumber\\
  &\left. \left\{ 1  +B \left(-\frac{2}{3}\sqrt{\pi}Y_{0}^{0} +\frac{8}{3}\sqrt{\frac{\pi}{5}}Y_{2}^{0}\right)  \right\} + Pe_{z}B   \sqrt{\frac{2\pi}{15}}\left(Y_{2}^{-1}+Y_{2}^{1} \right) m Y_{l}^{m}    \right. \nonumber \\
&\left.+\frac{Y_{l}^{m}}{\tau D_r}\left(1-\frac{A_{l}}{4\pi}   \right) +l(l+1)Y_{l}^{m}\right]=0,
\label{EQ:G0Eq2}
\end{align}
where $k_{lm}^{\prime}=\sqrt{l(l+1)-m(m+1)}$ and $k_{lm}^{\prime \prime}=\sqrt{l(l+1)-m(m-1)}$.

Now, multiplying the above equation with the conjugate spherical harmonic $Y_{r}^{s^{*}}(\mathbf{p})$, then integrating it over the unit sphere, and using the orthogonality of spherical harmonics, we obtain the following infinite sequence of linear equations for the $a_{l,m}$'s:

\begin{align}
\sum_{l=0}^{\infty}\sum_{m=-l}^{l}a_{l,m} &\Bigl\{ Pe_z\frac{\sqrt{\pi}(3-B)}{6}\left[ k_{lm}^{\prime} <r,s\vert Y_{0}^{0}\vert l,m+1 >-k_{lm}^{\prime \prime}<r,s \vert Y_{0}^{0} \vert l,m-1> \right] \nonumber\\
&+Pe_z\left(\frac{B}{3}\right) \sqrt{\frac{4\pi}{5}} \left[k_{lm}^{\prime}<r,s \vert Y_{2}^{0} \vert l,m+1 >-k_{lm}^{\prime \prime}<r,s \vert Y_{2}^{0} \vert l,m-1> \right] \nonumber \\
&+Pe_{z} B\sqrt{\frac{2\pi}{15}}\left[ (m-3)<r,s \vert Y_{2}^{-1} \vert l,m>+(m+3) <r,s \vert Y_{2}^{1} \vert l,m> \right] \nonumber\\
& +\delta_{l r}\delta_{m s}\left[\frac{1}{\tau D_r}\left(1-\frac{4\pi A_{l}}{(2l+1)} \right)+l(l+1) \right]\Bigr\}=0.
\label{EQ:eq_1}
\end{align}
In the above equation, the Clebsch-Gordan coefficients
\begin{equation*}
<r,s \vert Y_{a}^{b} \vert l,m>=\int d\mathbf{p} Y_{r}^{s^{*}}(\mathbf{p}) Y_{a}^{b}(\mathbf{p}) Y_{l}^{m}(\mathbf{p}),
\end{equation*}
will be evaluated using the Racah formula [\cite{arfken1999mathematical,doi1978dynamics,messiah1962quantum}]. 

Using $\Omega_0=G(\mathbf{p}) F(z)$ and (\ref{EQ:GpExp}) in the following normalization condition (see \textsection\ref{subsec:conc})
\begin{equation}
\int d \mathbf{p} \Omega=\int d \mathbf{p}(\Omega_{0}+\epsilon \Omega_{1}+\epsilon^{2}\Omega_{2}+\cdots)=F,
\label{EQ:eq_2}
\end{equation}
we get 
\begin{equation}
a_{0,0}=1/\sqrt{4\pi}.
\label{EQ:a00value}
\end{equation}
Substituting (\ref{EQ:a00value}), in (\ref{EQ:eq_1}), we have:
\begin{align}
\sum_{l=1}^{\infty}\sum_{m=-l}^{l}&a_{l,m}\Bigl\{ Pe_z\frac{\sqrt{\pi}(3-B)}{6}\left[ a_{lm}^{\prime} <r,s\vert Y_{0}^{0}\vert l,m+1 >-a_{lm}^{\prime \prime}<r,s \vert Y_{0}^{0} \vert l,m-1> \right] \nonumber\\
&+Pe_z\left(\frac{B}{3}\right) \sqrt{\frac{4\pi}{5}} \left[a_{lm}^{\prime}<r,s \vert Y_{2}^{0} \vert l,m+1 >-a_{lm}^{\prime \prime}<r,s \vert Y_{2}^{0} \vert l,m-1> \right] \nonumber \\
&+Pe_z B\sqrt{\frac{2\pi}{15}}\left[ (m-3)<r,s \vert Y_{2}^{-1} \vert l,m>+(m+3) <r,s \vert Y_{2}^{1} \vert l,m> \right] \nonumber \\
&+\delta_{l r}\delta_{m s}\left[ \frac{1}{\tau D_r}\left(1-\frac{A_{l}4\pi}{(2l+1)}\right)+ l(l+1)\right] \Bigr\} \nonumber\\
&=Pe_{z}B\sqrt{\frac{3}{10}}\left[<r,s \vert Y_{2}^{-1} \vert 0,0>-<r,s \vert Y_{2}^{1} \vert 0,0> \right].
\label{EQ:eq_3}
\end{align}
An appropriately truncated version of (\ref{EQ:eq_3}) is solved numerically to obtain the $a_{l,m}$'s, subject to convergence of the orientation distribution.

Now substituting $\Omega_0=G(\mathbf{p};Pe_z)F(z,t_2)$ in the right hand side of (\ref{EQ:Omega1}), after simplification, we obtain:
\begin{equation}
 Pe_z \nabla_{p}\cdot(\tilde{\dot{\mathbf{p}}}\Omega_{1})+\frac{1}{\tau D_{r}}\left[\Omega_{1}-\int d \mathbf{p}^{'} K(\mathbf{p}/\mathbf{p}^{'})\Omega_{1}(\mathbf{p}^{'}) \right]-\nabla^{2}_{p}\Omega_{1}=-\cos \theta G \frac{\partial F}{\partial z}-\cos \theta F\frac{\partial G}{\partial z}.\\
 \label{EQ:eq_4}
\end{equation}
 To evaluate $\partial G/\partial z$ in the equation above, we first differentiate (\ref{EQ:Omega0}) with respect to $z$:
\begin{equation}
Pe_z\ \nabla_{p}\cdot\left(\tilde{\dot{\mathbf{p}}} \frac{\partial G}{\partial z} \right)  +\frac{1}{\tau D_{r}}\left[\frac{\partial G}{\partial z}-\int d \mathbf{p}^{'} K(\mathbf{p}/\mathbf{p}^{'})\frac{\partial G(\mathbf{p}^{'})}{\partial z} \right]-\nabla^{2}_{p}\frac{\partial G}{\partial z} =- \frac{\partial Pe_{z}}{\partial z}\nabla_{p}\cdot(\tilde{\dot{\mathbf{p}}}G).
\label{EQ:GovEqG}
\end{equation}

Substituting for $G$ on the right hand side, using (\ref{EQ:GpExp}), and using the expansion
\begin{equation}
\frac{\partial G}{\partial z}=\sum_{l=0}^{\infty}\sum_{m=-l}^{l}b_{l,m}Y_{l}^{m}(\mathbf{p})
\label{EQ:DiffGexpansion}
\end{equation}
in (\ref{EQ:GovEqG}), we get:
\begin{align}
\sum_{l=0}^{\infty}\sum_{m=-l}^{l}b_{l,m}&\left[ Pe_z\ \nabla_{p}\cdot\left(\tilde{\dot{\mathbf{p}}} Y_{l}^{m} \right) +\frac{1}{\tau D_{r}}\left[Y_{l}^{m}-\int d \mathbf{p}^{\prime} K(\mathbf{p}/\mathbf{p}^{\prime})Y_{l}^{m}(\mathbf{p}^{\prime}) \right]  -\nabla^{2}_{p}Y_{l}^{m}\right] \nonumber \\
&=- Pe \frac{\partial \dot{\gamma}}{\partial z} \sum_{l^{\prime}=0}^{\infty}\sum_{m^{\prime} 
=-l^{\prime}}^{l^{\prime}} a_{l^{\prime},m^{\prime}}\nabla_{p}\cdot\left(\tilde{\dot{\mathbf{p}}}Y_{l^{\prime}}^{m^{\prime}}\right).
\label{EQ:blmeq_1}
\end{align}
Differentiating (\ref{EQ:eq_2}) with respect to $z$, using (\ref{EQ:DiffGexpansion}) and using the orthogonality of spherical harmonics, we get
\begin{equation}
b_{0,0}=0.
\label{EQ:b00}
\end{equation}
Now, following a procedure analogous to that leading to (\ref{EQ:eq_3}), the sequence of linear equations governing the $b_{l,m}$'s in (\ref{EQ:blmeq_1}) may be written as:
\begin{align}
\sum_{l=1}^{\infty}\sum_{m=-l}^{l}b_{l,m}& \Bigl\{ Pe_z \left( \frac{\sqrt{\pi}(3-B)}{6}\left[ k_{lm}^{\prime} <r,s\vert Y_{0}^{0}\vert l,m+1 >-k_{lm}^{\prime \prime}<r,s \vert Y_{0}^{0} \vert l,m-1> \right]\right. \nonumber\\
&\left. +\left(\frac{B}{3}\right) \sqrt{\frac{4\pi}{5}} \left[k_{lm}^{\prime}<r,s \vert Y_{2}^{0} \vert l,m+1 >-k_{lm}^{\prime \prime}<r,s \vert Y_{2}^{0} \vert l,m-1> \right] \right. \nonumber \\
& \left. +B\sqrt{\frac{2\pi}{15}}\left[ (m-3)<r,s \vert Y_{2}^{-1} \vert l,m>+(m+3) <r,s \vert Y_{2}^{1} \vert l,m> \right] \right) \nonumber \\
&+\delta_{l r}\delta_{m s}\left[\frac{1}{\tau D_r}\left(1-\frac{A_{l}4\pi}{(2l+1)}\right)+  l(l+1)\right] \Bigr\}\nonumber \\
&=-\sum_{l^{\prime}=0}^{\infty}\sum_{ l^{\prime} =-l^{\prime}}^{l} a_{l^{\prime},m^{\prime}}\Bigl\{ Pe\frac{\partial \dot{\gamma}}{\partial z} \left( \frac{\sqrt{\pi}(3-B)}{6}\left[ k_{l^{\prime}m^{\prime}}^{\prime} <r,s\vert Y_{0}^{0}\vert l^{\prime},m^{\prime}+1 > \right. \right. \nonumber \\
&\left. \left. -k_{l^{\prime}m^{\prime}}^{\prime \prime}<r,s \vert Y_{0}^{0} \vert l^{\prime},m^{\prime}-1> \right]+\left(\frac{B}{3}\right) \sqrt{\frac{4\pi}{5}} \left[k_{l^{\prime}m^{\prime}}^{\prime} <r,s \vert Y_{2}^{0} \vert l^{\prime},m^{\prime}+1 >  \right. \right.\nonumber \\
&\left. \left. -k_{l^{\prime}m^{\prime}}^{\prime \prime}<r,s \vert Y_{2}^{0} \vert l^{\prime},m^{\prime}-1> \right]+B\sqrt{\frac{2\pi}{15}}\left[ (m^{\prime}-3)<r,s \vert Y_{2}^{-1} \vert l^{\prime},m^{\prime}> \right. \right. \nonumber \\
 & \left. \left. +(m^{\prime}+3) <r,s \vert Y_{2}^{1} \vert l^{\prime},m^{\prime}> \right]\right)+\delta_{l^{\prime} r}\delta_{m^{\prime} s}  \left[\frac{1}{\tau D_r}\left(1-\frac{A_{l^{\prime}}4\pi}{(2l^{\prime}+1)}\right) l^{\prime}(l^{\prime}+1) \right] \Bigr\},
 \label{EQ:blmeq_2}
\end{align}
where, $k_{l^{\prime}m^{\prime}}^{\prime}=\sqrt{l^{\prime}(l^{\prime}+1)-m^{\prime}(m^{\prime}+1)}$ and $k_{l^{\prime}m^{\prime}}^{\prime \prime}=\sqrt{l^{\prime}(l^{\prime}+1)-m^{\prime}(m^{\prime}-1)}$. One can solve the system of linear equations (\ref{EQ:blmeq_2}) to determine $dG/dz$ as a truncated version of (\ref{EQ:DiffGexpansion}).

To solve for $\Omega_1$ in (\ref{EQ:Omega1}), on account of linearity, we first write:

\begin{equation}
\Omega_{1}=\Omega_{11}\frac{\partial F}{\partial z}+\Omega_{12} F
\label{EQ:Omega1Expansion}
\end{equation}
 and equating the coefficients of $\partial F/\partial z$ and $F$ on both sides, we get the following equations governing $\Omega_{11}$ and $\Omega_{12}$:
\begin{align}
 \label{EQ:Omega11} Pe_z \nabla_{p}\cdot(\tilde{\dot{\mathbf{p}}}\Omega_{11})   +\frac{1}{\tau D_{r}}\left[\Omega_{11}-\int d \mathbf{p}^{\prime} K(\mathbf{p}/\mathbf{p}^{\prime})\Omega_{11}(\mathbf{p}^{\prime}) \right] -\nabla^{2}_{p}\Omega_{11}&=-\cos \theta G, \\ 
 Pe_z \nabla_{p}\cdot(\tilde{\dot{\mathbf{p}}}\Omega_{12}) +\frac{1}{\tau D_{r}}\left[\Omega_{12}-\int d \mathbf{p}^{\prime} K(\mathbf{p}/\mathbf{p}^{\prime})\Omega_{12}(\mathbf{p}^{\prime}) \right]-\nabla^{2}_{p}\Omega_{12}&=-\cos \theta \frac{\partial G}{\partial z}.
\label{EQ:Omega12} 
\end{align}
Expanding
\begin{equation}
\Omega_{11}=\sum_{l=0}^{\infty}\sum_{m=-l}^{l}c_{l,m}Y_{l}^{m},
\label{EQ:Omega11Expansion}
\end{equation}
\begin{equation}
\Omega_{12}=\sum_{l=0}^{\infty}\sum_{m=-l}^{l}d_{l,m}Y_{l}^{m},
\label{EQ:Omega12Expansion}
\end{equation}
using (\ref{EQ:Omega1Expansion}) in (\ref{EQ:eq_2}) and using the orthogonality of spherical harmonics, we get
\begin{equation}
c_{0,0}=d_{0,0}=0.
\label{EQ:c00andd00}
\end{equation}
 From (\ref{EQ:GpExp}), (\ref{EQ:DiffGexpansion}), (\ref{EQ:eq_2}), (\ref{EQ:blmeq_2}), (\ref{EQ:Omega11Expansion}), (\ref{EQ:Omega12Expansion}), and following the numerical scheme adopted for solving (\ref{EQ:GovEqG}), we obtain the unknown coefficients $c_{l,m}$ and $d_{l,m}$.

Using (\ref{EQ:Omega1Expansion}), (\ref{EQ:Omega11Expansion}), (\ref{EQ:Omega12Expansion}) in  (\ref{EQ:Omega2}), integrating over the orientation degrees of freedom and using $\int d\mathbf{p} \Omega_{0}(\mathbf{p})=F $ (from (\ref{EQ:eq_2})),  we get the drift-diffusion equation describing the evolution of swimmer concentration profile
 \begin{equation}
\frac{\partial F}{\partial t_2}= \frac{\partial}{\partial z}\left(D_{zz}\frac{\partial F}{\partial z}-V_{z} F \right)
\label{NumericalEQ:DriftDiff_1}
 \end{equation}
 in terms of the slow time variable $t_2$, and with the drift and diffusivity coefficients being defined by:
 \begin{equation}
 D_{zz}=-\int d\mathbf{p} \cos \theta \Omega_{11}=-2\sqrt{\frac{\pi}{3}}c_{1,0}
 \label{EQ:DiffExpression}
 \end{equation}
 and
 \begin{equation}
 V_{z}=\int d\mathbf{p} \cos \theta \Omega_{12}=2\sqrt{\frac{\pi}{3}}d_{1,0}.
  \label{EQ:DriftExpression}
 \end{equation}
The drift-diffusion equation (\ref{NumericalEQ:DriftDiff_1}) appears as (\ref{EQ:DriftDiffEquation}) in \textsection\ref{subsec:conc}. The truncation in the above sequences of equations is such as to ensure converged values for the $D_{zz}$ and $V_z$. The truncated system has $l=L_{max}$. For  $\kappa=1.25,~ 2$, and 4, reported in the manuscript, we chose $L_{max}=40$; for the other aspect ratios, we chose $L_{max}=40$ and 80 for $Pe<200$ and $Pe\geq 200$, respectively.

We calculate the steady state swimmer concentration profile satisfying (\ref{NumericalEQ:DriftDiff_1}), while imposing the zero-flux condition at boundaries, which leads to
\begin{equation}
D_{zz}\frac{d F}{d z}-V_{z} F=0
\end{equation}
at $z=\pm 1$.

The solution of the above first order ordinary differential equation can be expressed as:
\begin{equation}
F(z)=\aleph \exp\left(\int_{-1}^{z} dz^{\prime} \frac{V_z(z^{\prime})}{D_{zz}(z^{\prime})}  \right),
\label{EQ:ConcProf}
\end{equation}
which appears as (\ref{EQ:SSConcProf}) in \textsection\ref{subsec:conc}. In (\ref{EQ:ConcProf}), the normalization constant $\aleph$ can be calculated while imposing the  condition$\int_{-1}^{1} d z F(z)=1$, which leads to:

\begin{equation}
\aleph=\frac{1} {\int_{-1}^{1} dz \exp\left(\int_{-1}^{z} dz^{\prime} \frac{V_z(z^{\prime})}{D_{zz}(z^{\prime})}  \right)  }.
\label{EQ:NormConst}
\end{equation}
We evaluate the integrals in (\ref{EQ:ConcProf}) and (\ref{EQ:NormConst}) using Simpson's rule.

\section{Small-$Pe$ expansion for the swimmer concentration profile}{\label{App:SmallPe}}

Herein, we obtain closed form analytical expression for the swimmer concentration, to O($Pe^2$). The orientation probability density at each order (in $\epsilon$) in the multiple scales analysis (see \textsection\ref{subsec:conc}) is further expanded in powers of $Pe$ to O($Pe^2$). Thus, the probability densities of O(1) ($G$ in (\ref{EQ:Omega0})), and at O($\epsilon$) ($\Omega_1$ in (\ref{EQ:Omega1})) are expanded to O($Pe^2$), with the imposition of the normalization condition: $\int d\textbf{p}\Omega=F(z)$. Here and in \textsection\ref{App:Near-Sphere}, we choose a coordinate system different from the one used in \textsection\ref{subsec:conc} (figure \ref{FIG:Schematic}), such that the polar ($\theta$) and azimuthal ($\phi$) angles  are measured from the negative vorticity and flow axes, respectively. We first consider the probability density at leading order in the multiple scales analysis, and expand it about the isotropic orientation distribution, for $Pe \ll 1$, as:
\begin{equation}
G=G_{0}+PeG_1+Pe^2G_{2}+\cdots
\label{EQ:GExp}
\end{equation}
with $G_0=1/4\pi$. 

At successive orders in $Pe$, we obtain:
\begin{eqnarray}
\label{EQ:G1} \mathcal{O}(Pe)&:& \frac{1}{\tau D_{r}}\left[G_{1}-\int d \mathbf{p}^{\prime} K(\mathbf{p}/\mathbf{p}^{\prime})G_{1}(\mathbf{p}^{\prime}) \right]-\nabla^{2}_{p}G_{1}=-\frac{\dot{\gamma}}{4\pi}\nabla_{p}\cdot( \mathbf{\tilde{\dot{p}}}),\\
\label{EQ:G2} \mathcal{O}(Pe^{2})&:& \frac{1}{\tau D_{r}}\left[G_{2}-\int d \mathbf{p}^{\prime} K(\mathbf{p}/\mathbf{p}^{\prime})G_{2}(\mathbf{p}^{\prime}) \right]-\nabla^{2}_{p}G_{2}=-\dot{\gamma}\nabla_{p}\cdot(G_{1} \mathbf{\tilde{\dot{p}}}).
\end{eqnarray}
 
We solve (\ref{EQ:G1}) for $G_1$, by using the modified Green's function $\mathcal{G}^{I}_{M}(\mathbf{p}/\mathbf{p{'}})$ [\cite{subramanian2009critical}] which satisfies:
\begin{equation}
\frac{1}{\tau D_{r}}\left[\mathcal{G}^{I}_{M}(\mathbf{p}/\mathbf{p^{\prime}})-\int d \mathbf{p}^{\prime\prime} K(\mathbf{p}/\mathbf{p}^{\prime\prime})\mathcal{G}^{I}_{M}(\mathbf{p}^{\prime\prime}/\mathbf{p}) \right]-\nabla^{2}_{p}\mathcal{G}^{I}_{M}(\mathbf{p}/\mathbf{p^{\prime}})=\delta(\mathbf{p}-\mathbf{p^{\prime}})-\frac{1}{4\pi},
\label{EQ:GreensFunc}
\end{equation}
where $\delta(\mathbf{p})$ denotes the Dirac delta function in orientation space, with the forcing in (\ref{EQ:GreensFunc}) being a localized source at an orientation $\mathbf{p}^{\prime}$, together with a compensating uniformly distributed sink over the remainder of the unit sphere. The modified Green's function result given in \cite{subramanian2009critical}, derived for the rotary diffusion of swimmers can be easily generalized to the case of run-and-tumble plus rotary diffusion, so as to arrive at the following expression for $\mathcal{G}^{I}_{M}(\mathbf{p}/\mathbf{p^{\prime}})$ as: 
\begin{equation}
 \mathcal{G}^{I}_{M}(\mathbf{p}/\mathbf{p^{\prime}})=\sum_{n=1}^{\infty}\sum_{m=-n}^{n} \frac{Y_{n}^{m^{*}}(\mathbf{p}^{\prime})}{D_n}Y_{n}^{m}(\mathbf{p}),
 \label{EQ:GreensFuncExp} 
 \end{equation}
where $D_{n}=\left[\frac{1}{\tau D_r}(1-\frac{4 \pi A_{n}}{2n+1})+n(n+1)\right].$

Using the expression for $\mathcal{G}^{I}_{M}(\mathbf{p}/\mathbf{p^{\prime}})$ in (\ref{EQ:GreensFuncExp}), the orientation probability density $G_1$, at O($Pe$), may be written as:
\begin{align}
G_1(\textbf{p})&=-\frac{\dot{\gamma}}{4\pi}\int d\mathbf{p}^{\prime} \mathcal{G}^{I}_{M}(\mathbf{p}/\mathbf{p}^{\prime}) \nabla_{p}\cdot( \mathbf{\tilde{\dot{p}}}(\mathbf{p}^{\prime}))\nonumber \\
&=\frac{3B\dot{\gamma}}{4\pi}\sum_{l=1}^{\infty}\sum_{m=-l}^{l} \frac{Y_{l}^{m}(\mathbf{p})}{D_{l}} \int d\mathbf{p}^{\prime} Y_{l}^{m^{*}}(\mathbf{p}^{\prime}) p_{1}^{\prime}p_{3}^{\prime}\nonumber\\
&=\frac{3i\dot{\gamma}B}{4\pi D_{2}}\sqrt{\frac{2\pi}{15}}(Y_{2}^{-2}-Y_{2}^{2}).
\label{SmallPeEQ:G1Expression}
\end{align}

Similarly, using the modified Green's function, the solution of (\ref{EQ:G2}) may also be expressed as:
\begin{equation}
G_2(\textbf{p})=-\dot{\gamma}\int d\mathbf{p}^{\prime}\mathcal{G}^{I}_{M}(\mathbf{p}/\mathbf{p}^{\prime}) \nabla_{p}\cdot(G_{1}(\mathbf{p}^{\prime}) \mathbf{\tilde{\dot{p}}(\mathbf{p}^{\prime})}).
\label{SmallPeEQ:G2Expression}
\end{equation}

Again, using (\ref{EQ:GreensFuncExp}), (\ref{SmallPeEQ:G1Expression}) in the above equation, expressing $\nabla_{p}\cdot(G_{1} \mathbf{\tilde{\dot{p}}})$ in terms of spherical harmonics, and using the orthogonality of the latter, we get:
\begin{equation}
G_2=\frac{\dot{\gamma}^{2}B}{2D^2_{2}}\sqrt{\frac{3}{10\pi}}(Y_{2}^{-2}+Y_{2}^{2})-\frac{3\dot{\gamma}^{2}B^2}{14D^2_{2}\sqrt{5\pi}}Y_{2}^{0}-\frac{\dot{\gamma}^{2}B^2}{2D_{2}D_{4}}\sqrt{\frac{5}{14\pi}}(Y_{4}^{-4}+Y_{4}^{4})+\frac{\dot{\gamma}^{2}B^2 }{14\sqrt{\pi}D_{2}D_{4}}Y_{4}^{0}.
\label{SmallPeEQ:G2Expression}
\end{equation}

From (\ref{EQ:GExp}), (\ref{SmallPeEQ:G1Expression}) and (\ref{SmallPeEQ:G2Expression}), we obtain the small $Pe$ of the leading order orientation probability density, $G(\mathbf{p})$, to O($Pe^2$) as:
\begin{align}
G&=\frac{1}{4\pi}+Pe\frac{3i\dot{\gamma}B}{4\pi D_{2}}\sqrt{\frac{2\pi}{15}}(Y_{2}^{-2}-Y_{2}^{2})\nonumber\\
&+Pe^2\left[\frac{\dot{\gamma}^{2}B}{2D^2_{2}}\sqrt{\frac{3}{10\pi}}(Y_{2}^{-2}+Y_{2}^{2})-\frac{3\dot{\gamma}^{2}B^2}{14D^2_{2}\sqrt{5\pi}}Y_{2}^{0}-\frac{\dot{\gamma}^{2}B^2}{2D_{2}D_{4}}\sqrt{\frac{5}{14\pi}}(Y_{4}^{-4}+Y_{4}^{4})+\frac{\dot{\gamma}^{2}B^2 }{14\sqrt{\pi}D_{2}D_{4}}Y_{4}^{0}\right].
\end{align}

Now, considering the equation at O($\epsilon$), given by (\ref{EQ:Omega1}), and expanding the unknown probability density as:
\begin{equation}
\Omega_{1}=\Omega_{10}+Pe\Omega_{11}+Pe^2\Omega_{12}+\cdots,
\label{SmallPeEQ:Omega1Expansion}
\end{equation}
leads to the following three equations at successive orders in $Pe$:
\begin{eqnarray}
\label{SmallPeEQ:Omega10GovEq}
\mathcal{O}(\epsilon)&:&\frac{1}{\tau D_{r}}\left[\Omega_{10}-\int d \mathbf{p}^{'} K(\mathbf{p}/\mathbf{p}^{'})\Omega_{10}(\mathbf{p}^{'}) \right]-\nabla^{2}_{p}\Omega_{10}=-\mathbf{p}\cdot\nabla_{\mathbf{x}}(FG_{0}), \\
\label{SmallPeEQ:Omega11GovEq} \mathcal{O}(\epsilon Pe)&:& \frac{1}{\tau D_{r}}\left[\Omega_{11}-\int d \mathbf{p}^{'} K(\mathbf{p}/\mathbf{p}^{'})\Omega_{11}(\mathbf{p}^{'}) \right]-\nabla^{2}_{p}\Omega_{11}=-\mathbf{p}\cdot\nabla_{\mathbf{x}}(F G_{1})-\nabla_{p}\cdot(\dot{\mathbf{p}}\Omega_{10}),\nonumber\\
\end{eqnarray}
\vspace{-.5cm}
\begin{equation}
\mathcal{O}(\epsilon Pe^{2}): \frac{1}{\tau D_{r}}\left[\Omega_{12}-\int d \mathbf{p}^{'} K(\mathbf{p}/\mathbf{p}^{'})\Omega_{12}(\mathbf{p}^{'}) \right]-\nabla^{2}_{p}\Omega_{12}=-\mathbf{p}\cdot\nabla_{\mathbf{x}}(F G_{2})-\nabla_{p}\cdot(\dot{\mathbf{p}}\Omega_{11}).
\label{SmallPeEQ:Omega12GovEq}
\end{equation}

The solutions of (\ref{SmallPeEQ:Omega10GovEq}) and (\ref{SmallPeEQ:Omega11GovEq}) may again be expressed as a convolution involving the modified Greens function. Using the modified Green's function (\ref{EQ:GreensFuncExp}) and $G_0=1/4 \pi$ in (\ref{SmallPeEQ:Omega10GovEq}), we have:
\begin{equation}
\Omega_{10}=-\frac{i}{4 \pi D_{1}}\sqrt{\frac{2\pi}{3}}(Y_{1}^{-1}+Y_{1}^{1})\frac{\partial F}{\partial z}.
\label{SmallPeEQ:Omega10Expression}
\end{equation}
Using (\ref{SmallPeEQ:G1Expression}) for $G_0$, (\ref{SmallPeEQ:Omega10Expression}) for $\Omega_{10}$, and the modified Green's function (\ref{EQ:GreensFuncExp}), the unknown probability density in (\ref{SmallPeEQ:Omega11GovEq}) can be expressed as:
\begin{eqnarray}
\Omega_{11}&=&\frac{\left[6D_{1}(-2F+\frac{\partial F}{\partial z}\dot{\gamma})B+\dot{\gamma} D_{2}\frac{\partial F}{\partial z}(5+3B)\right]}{20D^{2}_{1}D_{2}\sqrt{6\pi}}(Y_{1}^{-1}-Y_{1}^{1})\nonumber \\
&+&\frac{\left[ -6D_{1}F+(3D_{1}+4D_{2})\dot{\gamma}\frac{\partial F}{\partial z}  \right]B}{20D_{1}D_{2}D_{3}\sqrt{21\pi}}(Y_{3}^{-1}-Y_{3}^{1}) \nonumber \\
&+&\frac{\left[-6D_{1}F+(3D_{1}+4D_{2})\frac{\partial F}{\partial z}\dot{\gamma} \right]B}{4\sqrt{35\pi}D_{1}D_{2}D_{3}}(Y_{3}^{-3}-Y_{3}^{3}).
\label{SmallPeEQ:Omega11Expression}
\end{eqnarray}
Similarly, using (\ref{SmallPeEQ:G2Expression}) for $G_2$, (\ref{SmallPeEQ:Omega11Expression}) for $\Omega_{11}$ and the modified Green's function (\ref{EQ:GreensFuncExp}), the unknown probability density in (\ref{SmallPeEQ:Omega12GovEq}) can be expressed as:
\begin{align}
\Omega_{12}&= \frac{i\dot{\gamma}}{1400D^3_{1}D^{2}_{2}D_{3}\sqrt{6\pi}}\left[60BD^{2}_{1}D_{3}(4F-\frac{\partial F}{\partial z}\dot{\gamma})(-7+B) \right. \nonumber\\
& \left.+ B^{2}_{2}\frac{\partial F}{\partial z}\dot{\gamma}\left( 128 B^{2}B_{1}-7B_{3}(-25+9B^{2}) \right) \right. \nonumber \\
&\left.-6BD_{1}D_{2}(2F-\frac{\partial F}{\partial z})(16BD_{1}-7D_{3}(-5+3B))  \right](Y_{1}^{-1}+Y_{1}^{1})+(..)((Y_{3}^{-1}+Y_{3}^{1})) \nonumber\\
&+(...)(Y_{3}^{-3}+Y_{3}^{3})+(..)((Y_{5}^{-1}+Y_{5}^{1}))+(...)(Y_{5}^{-3}+Y_{5}^{3})+(...)(Y_{5}^{-5}+Y_{5}^{5}).
\label{SmallPeEQ:Omega12Expression}
\end{align}
In the above equation, we only mention the coefficients of $(Y_{1}^{-1}+Y_{1}^{1})$, since only these contribute to the drift, diffusivity, and hence the swimmer concentration profile (see (\ref{SmallPeEQ:DriftDiff_2}) below).

Considering (\ref{EQ:Omega2}), integrating over the orientation degrees of freedom and using the normalization condition $\int d\mathbf{p}\Omega_0=F$ (see (\ref{EQ:eq_2})), we get:
\begin{equation}
\frac{\partial F}{\partial t_{2}}=-\int d\mathbf{p} \frac{\partial \Omega_{1}}{\partial z}p_{3}.
\label{SmallPeEQ:DriftDiff_1} 
\end{equation}
Expressing $p_3$ in terms of spherical harmonics, we have:
\begin{equation}
\frac{\partial F}{\partial t_{2}}=-i\sqrt{\frac{2\pi}{3}}\int d\mathbf{p} (Y_{1}^{-1}+Y_{1}^{1}) \frac{\partial }{\partial z}\left(\Omega_{10}+Pe\Omega_{11}+Pe^{2}\Omega_{12}+\cdots\right).
\label{SmallPeEQ:DriftDiff_2} 
\end{equation}
From orthogonality, only the coefficients of $(Y_{1}^{-1}+Y_{1}^{1})$, in $\Omega_{10}, \Omega_{11}$ and $\Omega_{12}$, survive. Substituting for $\Omega_{10}, \Omega_{11}$, $\Omega_{12}$ from (\ref{SmallPeEQ:Omega10Expression}), (\ref{SmallPeEQ:Omega11Expression}), (\ref{SmallPeEQ:Omega12Expression}), the shear rate profile for the parabolic flow, $\dot{\gamma} = -2z$, and evaluating the orientation-space integrals, we get:

\begin{equation}
\frac{\partial F}{\partial t_{2}}=\frac{\partial}{\partial z}\left[ \left(D_{zz}^{(0)}+Pe^{2}D^{(2)}_{zz}+\cdots\right)\frac{\partial F}{\partial z}-\left(V^{(0)}_{z}+Pe^{2}V^{(2)}_{z}+\cdots\right)F \right],
\label{SmallPeEQ:DriftDiff_3}
\end{equation} 
where
\begin{align*}
D_{zz}^{(0)}&=\frac{1}{3D_{1}}, \\
D_{zz}^{(2)}&=\frac{\left[60D^{2}_{1}D_{3}(-7+B)B-6D_{1}D_{2}B\left(16D_{1}B-7D_{3}(-5+3B) \right) \right]z^2}{525D_{1}^{3}D_{2}^{2}D_{3}} \nonumber\\
&+\frac{\left[ D^{2}_{2}\left(-128B^{2}D_{1}+7D_{3}(-25+9B^{2}) \right)   \right]z^{2}}{525D_{1}^{3}D_{2}^{2}D_{3}},\\
V_{z}^{(0)}&=0, \\
V_{z}^{(2)}&=-\frac{2B\left[ 7D_{2}D_{3}(-5+3B)+4D_{1}\left(5D_{3}(-7+B)-4D_{2}B \right)  \right]z}{175D^{2}_{1}D_{2}^{2}D_{3}}.
\end{align*}
In (\ref{SmallPeEQ:DriftDiff_3}), the odd orders in $Pe$ don't contribute, as must be the case, so the concentration profile is invariant to flow reversal. We therefore expand the swimmer concentration as $F=F^{(0)}+Pe^{2}F^{(2)}+\cdots$. Substituting this expansion in (\ref{SmallPeEQ:DriftDiff_3}), 
imposing the zero-flux condition at the boundaries ($z=\pm 1$), we get:
\begin{equation}
D^{(0)}_{zz}\frac{{\partial F}^{(0)}}{\partial z}+Pe^{2}\left( D^{(0)}_{zz}\frac{{\partial F}^{(2)}}{\partial z}+D_{zz}^{(2)}\frac{\partial F^{(0)}}{\partial z}-V_{z}^{(2)}F^{(0)}
\right)+\cdots =0.
\label{SmallPeEQ:DriftDiff_5}
\end{equation}
At successive orders in $Pe$, we obtain:
\begin{align}
\label{SmallPeEQ:DriftDiff_61}O(1)&:D^{(0)}_{zz}\frac{{\partial F}^{(0)}}{\partial z}=0,\\
\label{SmallPeEQ:DriftDiff_62}O(Pe^2)&:D^{(0)}_{zz}\frac{{\partial F}^{(2)}}{\partial z}=-D_{zz}^{(2)}\frac{\partial F^{(0)}}{\partial z}+V_{z}^{(2)}F^{(0)}.
\end{align}
Using  (\ref{SmallPeEQ:DriftDiff_61}) in (\ref{SmallPeEQ:DriftDiff_62}), the leading order inhomogeneity (${F}^{(2)}$) is independent of the correction to the diffusivity and the O($Pe^2$) drift alone dictates the inhomogeneity. It is nevertheless worth noting that $D_{zz}^{(2)}$ exhibits a profile consistent with physical arguments - the fact that it is negative, and larger closer to the walls, where the transverse diffusivity is suppressed due to increasing flow alignment of the swimmer. Although not relevant to the steady state, $D_{zz}^{(2)}$ will influence the transient evolution from a generic initial condition.

Solving (\ref{SmallPeEQ:DriftDiff_61}) and (\ref{SmallPeEQ:DriftDiff_62}) for $F^{(0)}$ and $F^{(2)}$, while imposing the normalization condition of swimmer concentration $\int_{-1}^{1}dz F(z)=\int_{-1}^{1}dz (F^{(0)}+Pe^{2}F^{(2)}+\cdots)=1$, we get:
\begin{align}
F(z)=F^{(0)}+Pe^{2}F^{(2)}(z)+\cdots=\frac{1}{2}-Pe^{2}K_{2}(3z^{2}-1)-\cdots,
\end{align}
where 
\begin{equation}
K_2=\frac{B\left[7D_{2}D_{3}(-5+3B)+4D_{1}\left( 5D_{3}(-7+B)-4BD_{2}\right)   \right]}{350D_{1}D_{2}^{2}D_{3}}
\end{equation}
with $B$ as the Bretherton constant and $D_{i}$ being given by (\ref{EQ:GreensFuncExp}).
The comparison of the numerical and the analytical profiles of $F^{(2)}(z)$ for swimmers with $\kappa=15$ ($B=0.991$) and for different $Pe$ is given in figure \ref{FIG:1}a (inset) of \textsection\ref{Sec:3.1}.

\section{Concentration profile for the near-spherical swimmers}{\label{App:Near-Sphere}}

In the following, we obtain an analytical expression for the swimmer concentration profile, for an arbitrary $Pe$, but for near-spherical swimmers which correspond to an asymptotically small Bretherton constant ($B \ll 1$). As we did in \textsection\ref{App:SmallPe}, we start from the small-$\epsilon$ expansion of probability density ($\Omega$) in the multiple scales analysis, and expand the probability densities at each order in $\epsilon$, $G$ in (\ref{EQ:Omega0}) at O(1), and $\Omega_1$ in (\ref{EQ:Omega1}) at O($\epsilon$), to O($B$), while also imposing the normalization condition: $\int d\textbf{p}\Omega=F(z)$. To begin with, consider the governing equation (\ref{EQ:Omega0}) for $G$:
\begin{equation}
\frac{1}{\tau D_{r}}\left[G-\int d \mathbf{p}^{\prime} K(\mathbf{p}/\mathbf{p}^{\prime})G(\mathbf{p}^{\prime}) \right]-\nabla^{2}_{p}G+Pe_z\ \nabla_{p}\cdot(\mathbf{\dot{\tilde{p}}^{I}}G)=-Pe_z\ \nabla_{p}\cdot(\mathbf{\dot{\tilde{p}}^{II}}G),
\label{NearSphereEQ:G1m}
\end{equation}
where $\mathbf{\dot{\tilde{p}}^{I}}=\mathbf{W}\cdot \mathbf{p}/\dot{\gamma}$, $\mathbf{\dot{\tilde{p}}^{II}}=B[\mathbf{E}\cdot \mathbf{p}-\mathbf{p}(\mathbf{E}:\mathbf{p}\mathbf{p})]/\dot{\gamma}$, represent the rotation rates of the swimmer at O(1) and O($B$), respectively, due to the local ambient vorticity and rate of strain. In (\ref{NearSphereEQ:G1m}), for near-spherical swimmers, we expand $G$ in the parameter $B$, about the spatially homogeneous isotropically oriented base state
\begin{equation}
G=G_{0}+BG_{1}+\cdots.
\label{NearSphereEQ:GExp}
\end{equation}
 At leading order, one obtains the isotropically oriented state of spherical swimmers that spin at a uniform rate with the ambient vorticity. Thus, $G_0 = 1/4\pi$. At O($B$), we have the following governing equation for $G_1$:
\begin{equation}
  \frac{1}{\tau D_{r}}\left[G_{1}-\int d \mathbf{p}^{\prime} K(\mathbf{p}/\mathbf{p}^{\prime})G_{1}(\mathbf{p}^{\prime}) \right]-\nabla^{2}_{p}G_{1}+Pe_z\nabla_{p}\cdot(\mathbf{\dot{\tilde{p}}^{I}}G_1)=-Pe_z\nabla_{p}\cdot(\mathbf{\dot{\tilde{p}}^{II}}G_0).
  \label{NearSphereEQ:G1}
\end{equation}

To solve the above equation for $G_1$, we again seek a modified Green's function $\mathcal{G}^{II}_{M}(\mathbf{p}/\mathbf{p}^{\prime})$, that satisfies
\begin{align}
  \frac{1}{\tau D_{r}}\left[\mathcal{G}^{II}_{M}(\mathbf{p}/\mathbf{p}^{\prime})-\int d \mathbf{p}^{\prime\prime} K(\mathbf{p}/\mathbf{p}^{\prime\prime})\mathcal{G}^{II}_{M}(\mathbf{p}^{\prime\prime}/\mathbf{p}^{\prime}) \right]&-\nabla^{2}_{p}\mathcal{G}^{II}_{M}\left(\mathbf{p}/\mathbf{p}^{\prime}\right)+Pe_z\nabla_{p}\cdot\left(\mathbf{\dot{\tilde{p}}^{I}}\mathcal{G}^{II}_{M}(\mathbf{p}/\mathbf{p}^{\prime})\right)\nonumber\\
  &=\delta(\mathbf{p}-\mathbf{p}^{\prime})-\frac{1}{4\pi}.
  \label{NearSphereEQ:GreensFuncEquation}
\end{align}
Here, $\mathcal{G}^{II}_{M}(\mathbf{p}/\mathbf{p}^{\prime})$ differs from $\mathcal{G}^{I}_{M}(\mathbf{p}/\mathbf{p}^{\prime})$, given by (\ref{EQ:GreensFunc}), in that it includes the effect of vorticity-induced rotation in addition to the orientation decorrelation due to rotary diffusion and run-and-tumble dynamics. Writing down the modified Green's function as a series in spherical harmonics 
\begin{equation}
 \mathcal{G}^{II}_{M}(\mathbf{p}/\mathbf{p^{\prime}})=\sum_{n=0}^{\infty}\sum_{m=-n}^{n} c_{n}^{m}(\mathbf{p^{\prime}})Y_{n}^{m}(\mathbf{p}),
 \label{NearSphereEQ:GreensFunc} 
 \end{equation}
 substituting in (\ref{NearSphereEQ:GreensFuncEquation}), and following the procedure mentioned in appendix \ref{App:SmallPe}, we obtain 
 \begin{equation}
 c_{n}^{m}(\mathbf{p^{\prime}})=\frac{Y_{n}^{m^{*}}(\mathbf{p^{\prime}})}{R_{n, m}} \quad \text{for} \quad n \geq 1,
 \label{NearSphereEQ:GreensFuncCoeff}
 \end{equation}
 where $R_{n, m}=\left[\left(1-4\pi A_{n}/({2n+1})\right)/\tau D_r+n(n+1)- i mPe_z/2 \right].$ Note that the azimuthal degeneracy is broken by the presence of the $m$-dependent flow contribution. The effect of this flow contribution is to decrease the translational swimmer diffusivity at large $Pe$. For, large $Pe$, the spherical swimmer rotates with a period of O($D^{-1}_{r}Pe^{-1}$). The rotation and swimming lead to a mean free path of O($U_{s}Pe^{-1}D^{-1}_{r}$). Taken together with the decorrelation time of O($D_r^{-1}$), this implies a scaling of $(U_s^2/D_r)Pe^{-2}$ for the gradient component of the diffusivity. Thus, the rapid rotation induced by the ambient vorticity leads to an asymptotically small translational diffusivity at large $Pe$.
 
 Using the convolution integral involving the modified Green's function given by (\ref{NearSphereEQ:GreensFunc}), one may write the O($B$) swimmer probability density $G_1$, governed by (\ref{NearSphereEQ:G1}), in the following final form: 
 
 \begin{equation}
 G_1= -Pe_z\int d\mathbf{p}^{\prime}\mathcal{G}^{II}_{M}(\mathbf{p}/\mathbf{p}^{\prime})\nabla_{p}\cdot(\mathbf{\dot{\tilde{p}}^{II}}(\mathbf{p}^{\prime})G_0)  =\frac{3iPe_z}{4\pi}\sqrt{\frac{2\pi}{15}}\left(\frac{Y_{2}^{-2}}{R_{2,-2}}-\frac{Y_{2}^{2}}{R_{2,2}} \right).
 \label{NearSphereEQ:G1Expression}
 \end{equation}
 
Next, considering the swimmer probability density at O($\epsilon$) in the multiple scales analysis, and expanding in $B$:
\begin{equation}
\Omega_{1}=\Omega_{10}+B\Omega_{11}+\cdots,
\label{NearSphereEQ:Omega1Expansion}
\end{equation} 
 we obtain the following equations at successive orders in $B$: 
\begin{align}
   \label{NearSphereEQ:Omega10} O(1)&: \frac{1}{\tau D_{r}}\left[\Omega_{10}-\int d \mathbf{p}^{\prime} K(\mathbf{p}/\mathbf{p}^{\prime})\Omega_{10}(\mathbf{p}^{\prime}) \right]-\nabla^{2}_{p}\Omega_{10}+Pe_z\nabla_{p}\cdot(\mathbf{\dot{\tilde{p}}^{I}}\Omega_{10})=-(\mathbf{p}\cdot \nabla_{\mathbf{x}}F)G_{0}\\
    \label{NearSphereEQ:Omega11} O(B)&:  \frac{1}{\tau D_{r}}\left[\Omega_{11}-\int d \mathbf{p}^{\prime} K(\mathbf{p}/\mathbf{p}^{\prime})\Omega_{11}(\mathbf{p}^{\prime}) \right]-\nabla^{2}_{p}\Omega_{11}+Pe_z\nabla_{p}\cdot(\mathbf{\dot{\tilde{p}}^{I}}\Omega_{11})\nonumber\\
&\hspace*{6cm}=
-Pe_z\nabla_{p}\cdot(\mathbf{\dot{\tilde{p}}^{II}}\Omega_{10}) -(\mathbf{p}\cdot \nabla_{\mathbf{x}}F)G_{1}\nonumber\\
\end{align}

Substituting $G_0=1/4\pi$ and using (\ref{NearSphereEQ:GreensFuncEquation}), the solution of (\ref{NearSphereEQ:Omega10}) can be expressed as:
\begin{equation}
\Omega_{10}=-\frac{i}{4\pi}\sqrt{\frac{2\pi}{3}}\left(\frac{Y_{1}^{-1}}{R_{1,-1}}+\frac{Y_{1}^{1}}{R_{1,1}}  \right)\frac{\partial F}{\partial y}.
\label{NearSphereEQ:Omega10Expression}
\end{equation}

Similarly, from (\ref{NearSphereEQ:Omega10Expression}), (\ref{NearSphereEQ:G1Expression}) and (\ref{NearSphereEQ:GreensFuncEquation}), we obtain the solution of (\ref{NearSphereEQ:Omega11}) as:
\begin{align}
\Omega_{11}&=-\frac{Pe\sqrt{\frac{3}{2\pi}}\left[6F(4i+Pe_z)+5\frac{\partial F}{\partial z}\dot{\gamma}\left(-6i+Pe_z \right)    \right]}{5\left( 4i+Pe_z\right)\left( -6i+Pe_z\right)^{2}}\frac{Y_{1}^{-1}}{R_{1,-1}}\nonumber\\
&+\frac{Pe\sqrt{\frac{3}{2\pi}}\left[6F(-4i+Pe_z)+5\frac{\partial F}{\partial z}\dot{\gamma}\left(6i+Pe_z \right)    \right]}{5\left( -4i+Pe_z\right)\left( 6i+Pe_z\right)^{2}}\frac{Y_{1}^{1}}{R_{1,1}} \nonumber 
\\
&+(...)Y_{3}^{-3}+(...)Y_{3}^{3}+(...)Y_{3}^{1}+(...)Y_{3}^{-1}.
\label{NearSphereEQ:Omega11Expression}
\end{align}

As in \textsection\ref{App:SmallPe}, only the terms involving $(Y_{1}^{-1}+Y_{1}^{1})$ contribute to the diffusivity ($D_{zz}$) and drift ($V_z$). So for simplicity, we do not mention the coefficients of $Y_{3}^{\pm 1}, Y_{3}^{\pm 3}$ in (\ref{NearSphereEQ:Omega11Expression}). Integrating (\ref{EQ:Omega2}) over the orientation degrees of freedom and using $\int d\mathbf{p} G=1$ (from (\ref{EQ:eq_2})), we get:
\begin{equation}
\frac{\partial F}{\partial t_{2}}=-\int d \mathbf{p}p_3\frac{\partial\Omega_{1}}{\partial z}.
\label{NearSphereEQ:DriftDiff_1}
\end{equation}
Using (\ref{NearSphereEQ:Omega1Expansion}), (\ref{NearSphereEQ:Omega10Expression}), (\ref{NearSphereEQ:Omega11Expression}), in the above equation and integrating, gives:
\begin{equation}
\frac{\partial F}{\partial t_{2}}=\frac{\partial}{\partial z}\left[ \left(D^{(0)}_{zz}+BD^{(1)}_{zz}+\cdots \right)\frac{\partial F}{\partial z}-\left(V^{(0)}_{z}+BV^{(1)}_{z}+\cdots \right)F \right],
\label{NearSphereEQ:DriftDiff_2}
\end{equation}
where

\begin{align}
\label{NearSphereEQ:D0}D^{(0)}_{zz}&=\frac{4(72+2Pe^{2}_z)}{3(16+Pe^{2}_z)(36+Pe^{2}_{z})}, \\
\label{NearSphereEQ:D1}D^{(1)}_{zz}&=-\frac{4Pe^{2}_{z}}{(16+Pe^{2}_z)(36+Pe^{2}_{z})},\\
\label{NearSphereEQ:V0}V_{z}^{(0)}&=0, \\
\label{NearSphereEQ:V1}V_{z}^{(1)}&=\frac{24PePe_{z}(-84+Pe^{2}_{z})}{5(16+Pe^{2}_z)(36+Pe^{2}_{z})^2}.
\end{align}
In the above equation, $D^{(0)}_{zz}$ scales as O($1/Pe^2$) for large $Pe$, consistent with the scaling arguments above. From (\ref{NearSphereEQ:V1}), we see that the O($B$) drift at the wall, for near-spherical swimmers, changes sign when $Pe_z^2 = 84$. This may be rewritten as $\dot{\gamma}\vert_{z=-1}^2Pe^2=84$, whence, one obtains the threshold $Pe$ for drift reversal as: $Pe_{c}=\sqrt{21}\sim 4.5$. This critical $Pe$ is mentioned in \textsection\ref{subsec:PhaseDiag}, and appears in the $Pe-\kappa$ phase portrait in figure \ref{FIG:PhaseDiag}.

Now, to obtain the steady state swimmer concentration profile we expand 
\begin{equation}
F(z)=F^{(0)}+BF^{(1)}(z)+\cdots,
\label{NearSphereEQ:ConcProfExp}
\end{equation}
  substituting in the normalization condition $\int_{-1}^{1} dz F(z)=1,$ yields $F^{(0)}=1/2$ and
%

\begin{align}
\int_{-1}^{1} dz F^{(1)}(z)=0.
\label{NearSphereEQ:F1Eq}
\end{align}

Using $F^{(0)}=1/2$, substituting (\ref{NearSphereEQ:ConcProfExp}) in (\ref{NearSphereEQ:DriftDiff_2}),  and equating the coefficients of $B$ on both sides, we get:
\begin{equation}
\frac{d F^{(1)}}{dz}=-\frac{V^{(1)}_{z}}{2D_{zz}^{(0)}} .
\label{NearSphereEQ:DiffDriftEq_1}
\end{equation}

From (\ref{NearSphereEQ:D0}), (\ref{NearSphereEQ:V1}) and (\ref{NearSphereEQ:F1Eq}), 
the solution of (\ref{NearSphereEQ:DiffDriftEq_1}), is:

\begin{align}
F^{(1)}(z)=\frac{9}{40} \left[-2 - \frac{30}{\left(9 + Pe^{2}z^2 \right)}+ \frac{(16 \tan^{-1}(Pe/3))}{Pe} + \log{(9 + Pe^2)} -\log{\left(9 + Pe^{2}z^2\right)}\right]
\label{NearSphereEQ:ConcProfF1}
\end{align}
 Thus from (\ref{NearSphereEQ:ConcProfExp}), (\ref{NearSphereEQ:ConcProfF1}) and using $F^{(0)}=1/2$, we obtain the following steady state concentration profile for near-spherical swimmers:
  
\begin{align}
F(z)&=\frac{1}{2}+B\frac{9}{40} \left[-2 - \frac{30}{\left(9 + Pe^{2}z^2 \right)}+ \frac{(16 \tan^{-1}(Pe/3))}{Pe} + \log{(9 + Pe^2)} -\log{\left(9 + Pe^{2}z^2\right)}\right].
\label{NearSphereEQ:ConcProfile}
\end{align}  

\begin{figure}
\includegraphics[width=.9\textwidth]{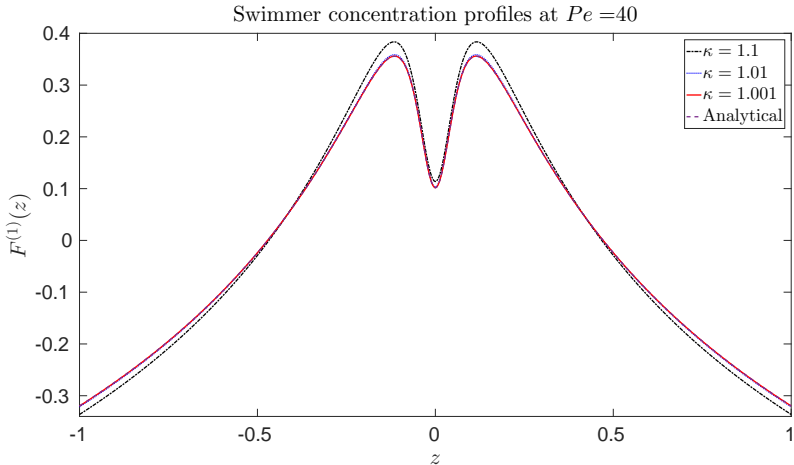}
\caption{Comparison of $F^{(1)}(z)=(F-F^{(0)})/B$, obtained from the numerics and the analytical solution ((\ref{NearSphereEQ:ConcProfF1})) for near-spherical swimmers at $Pe=40$.}
\label{AppD_FIG:Fig2}
\end{figure}

In figure \ref{AppD_FIG:Fig2}, we compare the numerical and analytical results of $F^{(1)}(z)$ for near-spherical swimmers at $Pe=40$. Expectedly, the numerical profiles approach the analytical one for $\kappa \rightarrow 1$. More importantly, the spatial structure of the swimmer concentration profile is similar to that obtained for finite $B$- a pair of concentration maxima on either side of the channel centerline. 

\end{appendix}

\bibliography{JFMReferences}
\bibliographystyle{jfm}

\end{document}